\documentclass[a4paper,11pt]{article}

\usepackage{jheppub} 
\usepackage[T1]{fontenc} 

\usepackage{graphicx,multirow,subfigure}
\usepackage{float}
\usepackage{bm}
\usepackage{amsmath}
\usepackage{amssymb}
\usepackage{amscd}
\usepackage{latexsym}
\usepackage{slashed}
\usepackage{color}
\usepackage{graphicx}
\usepackage{ulem}
\usepackage{color}
\usepackage{verbatim}
\usepackage{rotating}
\usepackage{diagbox}

\def\bea{\begin{eqnarray}}
\def\eea{\end{eqnarray}}

\title{ \centering 
Novel SM-like Higgs decay into displaced heavy neutrino pairs in $U(1)'$ models
}

\author[a]{Elena Accomando,}
\author[a,b]{Luigi Delle Rose,}
\author[a,b]{Stefano Moretti,} 
\author[b]{Emmanuel Olaiya,}
\author[b]{Claire H. Shepherd-Themistocleous}

\emailAdd{E.Accomando@soton.ac.uk}
\emailAdd{L.Delle-Rose@soton.ac.uk}
\emailAdd{S.Moretti@soton.ac.uk}
\emailAdd{Emmanuel.Olaiya@stfc.ac.uk}
\emailAdd{Claire.Shepherd@stfc.ac.uk}

\affiliation[a]{School of Physics and Astronomy, University of Southampton, Highfield, Southampton SO17 1BJ, United Kingdom}
\affiliation[b]{Particle Physics Department, Rutherford Appleton Laboratory, Chilton, Didcot, Oxon OX11 0QX, United Kingdom}

\abstract{\noindent
We examine the observability of heavy neutrino ($\nu_h$) signatures of a $U(1)'$ 
enlarged Standard Model (SM) encompassing three heavy Majorana
neutrinos alongside the known light neutrino states at the the Large Hadron Collider (LHC). We show that heavy neutrinos can be rather long-lived particles
producing distinctive displaced vertices that can be accessed in the CERN LHC detectors.  
We concentrate here on the gluon fusion production mechanism $gg\to H_{1,2}\to \nu_h\nu_h$, where $H_1$ is the discovered SM-like Higgs and $H_2$ is a heavier state, yielding  displaced leptons following $\nu_h$ decays into weak gauge bosons. Using data collected by the end of the LHC Run 2, these signatures would prove to be accessible with negligibly small background.}

\begin{document}
\maketitle
\flushbottom

\section{Introduction}
\label{sec:introduction}

The evolution of the three Standard Model (SM) gauge couplings through the Renormalisation Group Equations (RGEs) shows a remarkable convergence, although only approximate, around $10^{15}$ GeV (this feature is even more evident in a Supersymmetric context, where a near perfect convergence is achieved in presence of light sparticle states). This represents one of the largest hints in favour of a Grand Unification Theory (GUT) which embeds the SM symmetry group into a larger gauge structure. One of the main predictions of a GUT is the appearance of an extra $U(1)'$ gauge symmetry which can be broken at energies accessible at the CERN Large Hadron Collider (LHC). 
There are several realisations of GUTs that can predict this, such as $E_6$, String Theory motivated, $SO(10)$ and Left-Right (LR) symmetric models \cite{Langacker:1980js,Hewett:1988xc,Faraggi:1990ita,Faraggi:2015iaa,Faraggi:2016xnm,Randall:1999ee,Accomando:2010fz}, for example. 

At the Electro-Weak (EW) scale these Abelian extensions of the SM, in which the gauge group is enlarged by an extra $U(1)'$ symmetry, are also characterised by a new scalar field, heavier than the SM-like Higgs. The Vacuum Expectation Value (VEV) of this new scalar field can lie in the TeV range providing the mass for an additional heavy neutral gauge boson, $Z'$, associated to the spontaneous breaking of $U(1)'$. In this case an enlarged flavour sector is also always present. Indeed, in this class of models, the cancellation of the $U(1)'$ gauge and gravitational anomalies naturally predicts Right-Handed (RH) neutrinos at the TeV scale which realise a low-scale seesaw mechanism. Finally, a suitable
$U(1)'$ symmetry and the scale of its breaking can be tightly connected to
the baryogenesis mechanism through leptogenesis.

The special case in which the conserved charge of the extra Abelian symmetry is the $B-L$ number, with $B$ and $L$ the baryon and lepton charges, respectively, is particularly attractive from a phenomenological point of view. The minimal $B-L$ low-energy extension of the SM, consisting of a further $U(1)_{B-L}$ gauge group, predicts: three heavy RH neutrinos, one extra heavy neutral gauge boson, $Z'$, and an additional Higgs boson generated through the $U(1)_{B-L}$ symmetry breaking. This model has potentially interesting signatures at
hadron colliders, particularly the LHC. 
Those signatures pertaining to the $Z'$ and the (enlarged) Higgs sector have been extensively
studied in literature \cite{Khalil:2007dr,Basso:2008iv,Basso:2010hk,Basso:2010pe,Basso:2010yz,Basso:2010jm,Accomando:2010fz,Basso:2011na,Basso:2012sz,Basso:2012ux,Accomando:2013sfa,Accomando:2015cfa,Accomando:2015ava, Okada:2016gsh}. In this paper, we look at the heavy neutrino sector of a minimal Abelian extension of the SM. In fact, after the diagonalisation of the neutrino mass matrix realising the seesaw mechanism, we obtain three very light Left-Handed (LH) neutrinos ($\nu_l$), which are identified as SM neutrinos, and three heavy RH neutrinos ($\nu_h$). The latter have an extremely small mixing with the light $\nu_l$'s thereby providing very small but non-vanishing couplings to the gauge bosons. Moreover, owing to the mixing in the scalar sector, the Yukawa interaction of the heavy neutrinos with the heavier Higgs, $H_2$, also provides the coupling of the $\nu_h$'s to the SM-like Higgs boson, $H_1$. These non-zero couplings enable, in particular, the $gg\to H_1\to \nu_h\nu_h$ production mode if $m_{\nu_h} < M_{H_1}/2$ and the subsequent $\nu_h \rightarrow l^\pm W^{\mp *}$, $\nu_h \rightarrow \nu_l Z^{*}$ (off-shell) decays. The novelty of this signature compared to the existing literature is that the SM-like Higgs decays preferably into heavy RH neutrino pairs. A number of recent studies emphasize decay channels where the SM-like Higgs decays instead into one light and one heavy neutrino (see for example Refs.~\cite{Gago:2015vma,BhupalDev:2012zg,Cely:2012bz,Shoemaker:2010fg,Antusch:2016vyf}). 

The signature discussed in this paper gives rise to a different final state, as compared to this existing literature, which require dedicated experimental strategies to be detected.
Preliminary studies were performed in Ref.~\cite{Brooijmans:2012yi}, at parton level. In this paper, we refine the analysis in a more realistic way, taking into account full detector simulation. 

The heavy neutrino couplings to the weak gauge bosons ($W^\pm$ and $Z$) are
proportional to the ratio of light and heavy neutrino masses, which is extremely small.  
Therefore the decay width of the heavy neutrino is small and its lifetime large. The heavy neutrino can therefore be a long-lived particle and, over a large portion of the $U(1)'$ parameter space, its lifetime can be such that it can decay inside the LHC detectors, thereby producing a very distinctive signature with Displaced Vertices (DVs). 

In the case of a DV associated to a leptonic track (particularly a muon hit in the muon chamber), this represents a signal with a negligibly small background contribution (the background contribution is somewhat greater for an electron track stemming in the Electro-Magnetic (EM) calorimeter). In fact, for sufficiently large lifetimes, even jet decays of the heavy neutrino populating the hadronic calorimeter could be distinguished from those induced by a $B$-hadron\footnote{An experimentally resolvable non-zero lifetime along with a mass determination for the heavy neutrino would potentially also enable a determination of the light neutrino mass, as remarked in Ref.~\cite{Basso:2008iv}.}.

It is the purpose of this paper to examine the aforementioned heavy neutrino production and decay phenomenology at the LHC by firstly establishing the regions of the $U(1)'$ parameter space which have survived current theoretical and experimental constraints as well as eventually assessing the future scope of the LHC by accessing these signatures with DVs.

This paper is organised as follows. Sect.~\ref{sec:model} reviews
the model under study together with an overview of its allowed parameter
space. Sect.~\ref{sec:ProdDecay} illustrates the production and decay phenomenology of heavy neutrinos while
 Sect.~\ref{sec:eventsimulation} presents our Monte Carlo (MC) analysis. 
Finally, after a few remarks on the case of the heavy Higgs mediation (Sect.~\ref{sec:H2}),  
Sect.~\ref{sec:summa} concludes.

\section{The model}
\label{sec:model}
We study a minimal renormalizable Abelian extension of the SM with  only the  matter content necessary to satisfy the cancellation of the gauge and the gravitational anomalies. In this respect, we augment each of the three lepton families by a RH neutrino which is a singlet under the SM gauge group with $B-L$ = $-1$ charge.
In the scalar sector we introduce a complex scalar field $\chi$, besides the SM-like Higgs doublet $H$, to trigger the spontaneous symmetry breaking of the extra Abelian gauge group. The new scalar $\chi$ has $B-L$ = 2 charge and is a SM singlet. Its VEV, $x$, gives mass to the new heavy neutral gauge boson $Z'$ and provides the Majorana mass to the RH neutrinos through a Yukawa coupling. The latter dynamically implements the type-I seesaw mechanism. 

The presence of two Abelian gauge groups allows for a gauge invariant kinetic mixing operator of the corresponding Abelian field strengths. For the sake of simplicity, the mixing is removed from the kinetic Lagrangian through a suitable transformation (rotation and rescaling), thus restoring its canonical form. It is, therefore, reintroduced, through the coupling $\tilde g$, in the gauge covariant derivative which thus acquires a non-diagonal structure
\bea
\label{eq:gaugecovder}
\mathcal D_\mu = \partial_\mu + \ldots + i g_1 Y B_\mu + i \left( \tilde g Y + g'_1 Y_{B-L} \right) B_\mu',
\eea
where $Y$ and $Y_{B-L}$ are  the hypercharge and the $B-L$ quantum numbers, respectively, while $B_\mu$ and $B'_\mu$ are the corresponding Abelian fields. Other parameterisations, with a non-canonical diagonalised kinetic Lagrangian and a diagonal covariant derivative, are, however, completely equivalent.
The details of the kinetic mixing and its relation to the $Z-Z'$ mixing, which we omit in this work, can be found in \cite{Basso:2008iv,Coriano:2015sea,Accomando:2016sge}. 
Here we comment on some of the features of the model. 

The $Z-Z'$ mixing angle in the neutral gauge sector is strongly bounded indirectly by the EW Precision Tests (EWPTs) and directly by the LHC data \cite{Langacker:2008yv,Erler:2009jh,Cacciapaglia:2006pk,Salvioni:2009mt,Accomando:2016sge} to small values $|\theta'| \lesssim 10^{-3}$. In the $B-L$ model under study, we find
\bea
\label{eq:thetapexpandend}
\theta ' \simeq \tilde g \frac{M_Z \, v/2}{M_{Z'}^2 - M_Z^2} \,,
\eea
where $v$ is the VEV of the SM-like Higgs doublet, $H$. In this case, the bound on the $Z-Z'$ mixing angle can be satisfied by either $\tilde g \ll 1$ or $M_Z / M_{Z'} \ll 1$, the latter allowing for a generous interval of allowed values for $\tilde g$. 

\begin{figure}
\centering
\includegraphics[scale=0.7]{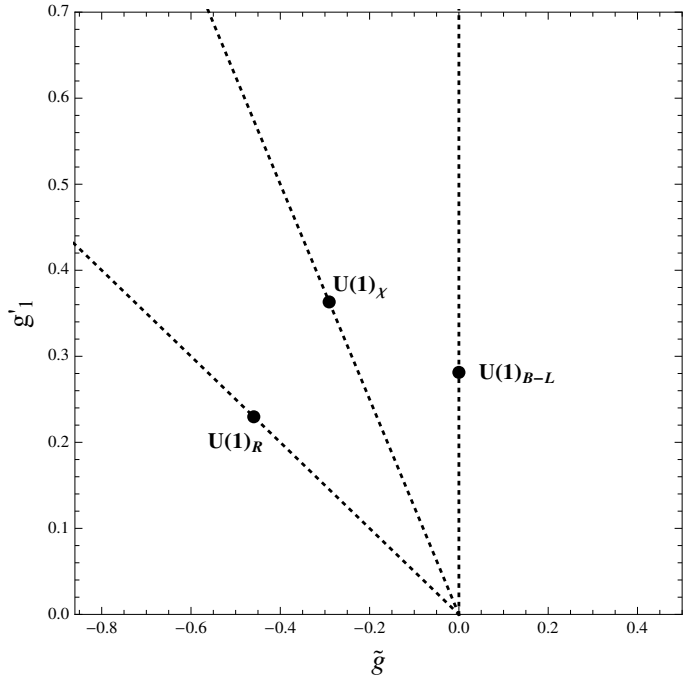}
\caption{Three typical $U(1)'$ charge assignments identified by the dashed lines in the $\tilde g - g'_1$ plane. The dots represent particular benchmark models described in the literature.
\label{Fig.Models}}
\end{figure}

It is also worth mentioning that a continuous variation of $\tilde g$ spans over an entire class of anomaly-free Abelian extensions of the SM with three RH neutrinos. 
Specific models, often studied in the literature, are identified by a particular choice of the two gauge couplings $g'_1$ and $\tilde g$. For instance, one can recover the pure $B-L$ model by setting $\tilde g = 0$. This choice corresponds to the absence of $Z-Z'$ mixing at the EW scale. Analogously, the Sequential SM is obtained for $g'_1 = 0$, the $U(1)_R$ extension is described by the relation $\tilde g = - 2 g'_1$, while the $U(1)_\chi$, generated at low scale in the $SO(10)$ scenario, is realised by $\tilde g = - 4/5 g'_1$. These models are graphically displayed in the plane $(\tilde g, g'_1$) in Fig.~\ref{Fig.Models}.

Moreover, there is no loss of generality in choosing the $U(1)_{B-L}$ as a reference gauge symmetry because any arbitrary $U(1)'$ gauge group can always be recast into a linear combination of $U(1)_Y$ and $U(1)_{B-L}$.

After spontaneous symmetry breaking, two mass eigenstates, $H_{1,2}$, with masses $m_{H_{1,2}}$, are obtained from the orthogonal transformation of the neutral components of $H$ and $\chi$. The mixing angle of the two scalars is denoted by $\alpha$. Moreover, we choose $m_{H_1} \le m_{H_2}$ and we identify $H_1$ with the $125$ GeV SM-like Higgs discovered at the CERN LHC.
The couplings between the light (heavy) scalar and the SM particles are equal to the SM rescaled by $\cos \alpha$ ($\sin \alpha$). The interaction of the light (heavy) scalar with the extra states introduced by the Abelian extension, namely the $Z'$ and the heavy neutrinos, is instead controlled by the complementary angle $\sin \alpha$ ($\cos \alpha$).

Finally, the Yukawa Lagrangian is
\bea
\mathcal L_Y =  \mathcal L_Y^{SM}  - Y_\nu^{ij} \, \overline{L^i} \, \tilde H \, {\nu_R^j}   - Y_N^{ij} \, \overline{(\nu_R^i)^c} \, {\nu_R^j} \, \chi + \, h.c., 
\eea
where $\mathcal L_Y^{SM}$ is the SM contribution. The Dirac mass, $m_D = 1/\sqrt{2}\, v Y_\nu$, and the Majorana mass for the RH neutrinos, $M = \sqrt{2}\, x Y_N$, are dynamically generated through the spontaneous symmetry breaking and, therefore, the type-I seesaw mechanism is automatically realised. Notice that $M$ can always be taken real and diagonal without loss of generality. For $M \gg m_D$, the masses of the physical eigenstates, the light and the heavy neutrinos, are, respectively, given by $m_{\nu_l} \simeq - m_D^T  M^{-1} m_D$ and $m_{\nu_h} \simeq M$. The light neutrinos are dominated by the LH SM components with a very small contamination of the RH neutrinos, while the heavier ones are mostly RH. The contribution of the RH components to the light states is proportional to the ratio of the Dirac and Majorana masses. After rotation into the mass eigenstates the charged and neutral currents interactions involving one heavy neutrino are given by
\bea
\mathcal L = \frac{g_2}{\sqrt{2}} \, V_{\alpha i} \, \bar l_\alpha \gamma^\mu P_L \nu_{h_i} \, W^-_\mu + \frac{g_Z}{2 \cos \theta_W} V_{\alpha \beta} V_{\alpha i}^* \, \bar \nu_{h_i} \gamma^\mu P_L \nu_{l_\beta} \, Z_\mu
\eea
where $\alpha, \beta = 1,2,3$ for the light neutrino components and $i = 1,2,3$ for the heavy ones. The sum over repeated indices is understood. In particular $V_{\alpha \beta}$ corresponds to the Pontecorvo-Maki-Nakagawa-Sakata (PMNS) matrix while $V_{\alpha i}$ describes the suppressed mixing between light and heavy states. Notice also that the $Z \nu_h \nu_h$ vertex is $\sim V_{\alpha i}^2$ and, therefore, highly dumped. These interactions are typical of a type-I seesaw extension of the SM. The existence of a scalar field generating the Majorana mass for RH neutrinos through a Yukawa coupling, which is a characteristic feature of the Abelian extensions of the SM, allows for a new and interesting possibility of producing a heavy neutrino pair from the SM-like Higgs (besides the obvious heavy Higgs mode). The corresponding interaction Lagrangian is given by 
\bea
\mathcal L = - \frac{1}{\sqrt{2}}Y_N^{k} \sin \alpha \, H_1 \, \bar \nu_{h_k} \nu_{h_k} = - g'_1 \frac{m_{\nu_{h,k}}}{M_{Z'}}  \sin \alpha \, H_1 \, \bar \nu_{h_k} \nu_{h_k},
\eea
where, in the last equation, we have used $x \simeq M_{Z'}/(2 g'_1)$. This expression for the VEV of $H_2$, $x$, neglects the sub-leading part that is proportional to $\tilde g$. For our purposes, this approximation can be safely adopted \cite{Basso:2010jm}. The interaction between the light SM-like Higgs and the heavy neutrinos is not suppressed by the mixing angle $V_{\alpha i}$ but is controlled by the Yukawa coupling $Y_N$ and the scalar mixing angle $\alpha$. 

For illustrative purposes we assume that the PMNS matrix is equal to the identity matrix and that both neutrino masses, light and heavy, are degenerate in  flavour. In this case the elements of the neutrino mixing matrix $V_{\alpha i}$ are simply given by $m_D/M \simeq \sqrt{m_{\nu_l}/m_{\nu_h}}$.

\section{Production and decay}
\label{sec:ProdDecay}
In this section, we focus on the production of heavy neutrino pairs coming from the decay of the light SM-like Higgs, $H_1$, at the LHC. The corresponding cross section can be written as
\bea
\label{eq:sigmaxBR}
\sigma(pp \rightarrow H_1 \rightarrow \nu_h \nu_h) = \cos^2 \alpha \, \sigma(pp \rightarrow H_1)_\textrm{SM} \frac{\Gamma( H_1 \rightarrow \nu_h \nu_h)}{\cos^2 \alpha \, \Gamma^\textrm{tot}_\textrm{SM} + \Gamma( H_1 \rightarrow \nu_h \nu_h)} \,,
\eea
where $\sigma(pp \rightarrow H_1)_\textrm{SM}$ and $\Gamma^\textrm{tot}_\textrm{SM}$ are the production cross section and total decay width of the SM Higgs state, respectively, while $\Gamma( H_1 \rightarrow \nu_h \nu_h)$ is the partial decay width of the SM-like $H_1$ boson  into two heavy neutrinos (summed over the three families). The partial decay width which reads as
\bea
\Gamma( H_1 \rightarrow \nu_h \nu_h) = \frac{3}{2}  \frac{m_{\nu_h}^2}{x^2} \sin^2 \alpha \frac{m_{H_1}}{8 \pi} \left( 1 -  \frac{4 m_{\nu_h}^2}{m_{H_1}^2} \right)^{3/2},
\eea
where we have safely neglected the contribution proportional to the neutrino Dirac mass. As intimated, 
it can be seen that the $H_1$ production cross section scales with $\cos^2\alpha$ with respect to the SM, where the SM is recovered for $\alpha = 0$, in which case we have $\sigma(pp \rightarrow H_1)_\textrm{SM} = 43.92$ pb (gluon channel) \cite{LHCHXSWG}. The total width of the SM Higgs is $\Gamma^\textrm{tot}_\textrm{SM} = 4.20 \times 10^{-3}$ GeV \cite{Heinemeyer:2013tqa}. 

\begin{figure}
\centering
\includegraphics[scale=0.8]{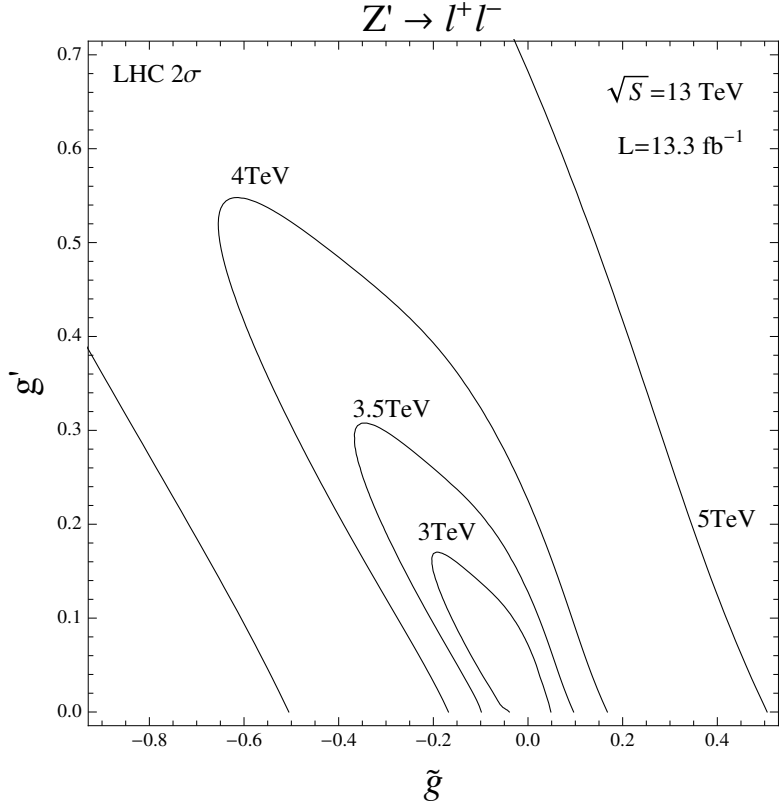}
\caption{Significance analysis for the di-lepton ($l = e, \mu$) DY channel at the LHC for different $Z'$-boson masses. Acceptance cuts for the electron and muon channels are applied and the corresponding efficiency factors are included. The two channels are then combined and the limits are extracted. \label{Fig.DrellYan}}
\end{figure}

\begin{figure}
\centering
\subfigure[]{\includegraphics[scale=0.79]{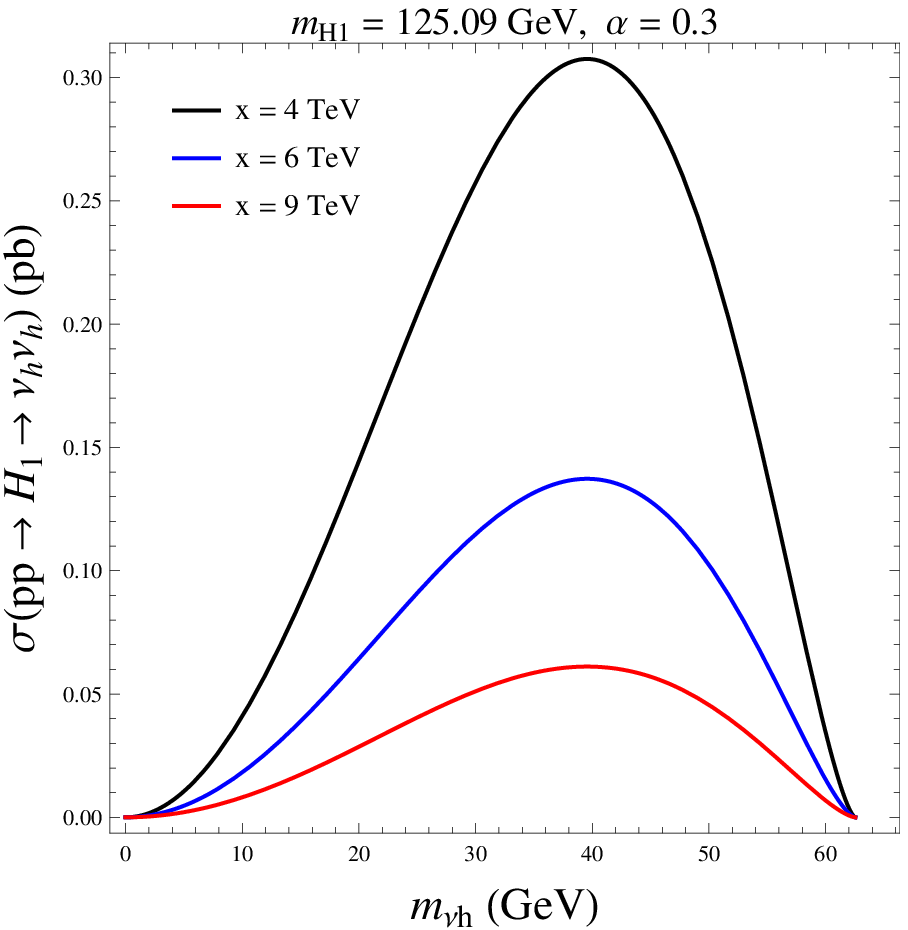}} \quad
\subfigure[]{\includegraphics[scale=0.79]{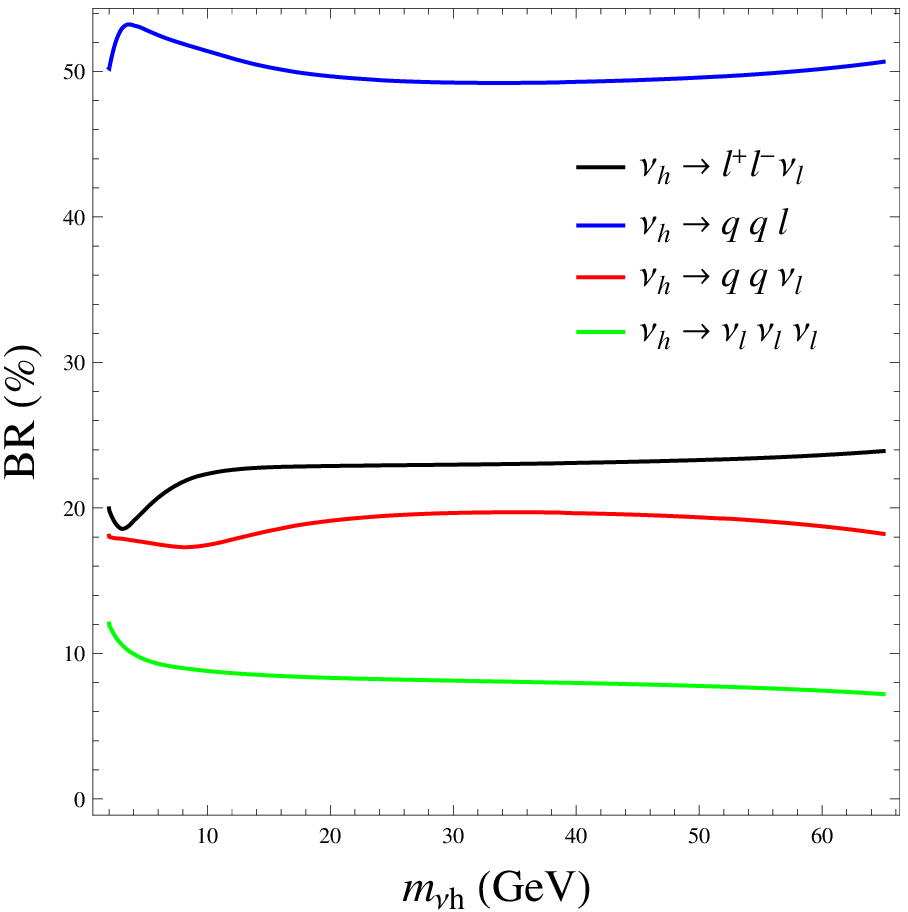}}
\caption{(a) $\sigma \times \textrm{BR}$ for the process $pp \rightarrow H_1 \rightarrow \nu_h \nu_h$ at the LHC, for 13 TeV proton-proton collisions as a function of the heavy RH neutrino mass. 
From top to bottom, the three curves display the cross section for three different values of the VEV of the heavier Higgs: $x$ = 4, 6, 9 TeV. The Higgs mixing angle is set to be $\alpha$ = 0.3. (b) Heavy neutrino branching ratios as a function of its mass. \label{Fig.SigmaBr}}
\end{figure}

To a large extent, the cross section in Eq.~(\ref{eq:sigmaxBR}) depends upon three parameters, namely, the mixing angle in the scalar sector, $\alpha$, the mass of the heavy neutrinos, $m_{\nu_h}$, and the VEV of the extra scalar, $x$ (or, equivalently, the Yukawa coupling $Y_N$). The dependence on $\tilde g$ is in fact negligible. The processes we are considering and the expression of the corresponding $\sigma$ remain unaffected in every extension of the SM in which the Majorana mass of the heavy neutrinos is dynamically generated by a SM-singlet scalar field sharing a non-zero mixing with the SM-like Higgs doublet. This scenario is naturally realised in the $U(1)'$ extension of the SM in which, we recall, the VEV $x$ is related to the mass of the $Z'$ through $x = M_{Z'}/(2 g')$. These parameters are obviously constrained by the $Z'$-boson direct search at the LHC. For this reason, in Fig.~\ref{Fig.DrellYan}, we show the exclusion limits at 95\% Confidence Level (CL) that have been extracted from the Drell-Yan (DY) analysis at the LHC with $\sqrt{S} = 13$ TeV and $\mathcal L = 13.3$ fb$^{-1}$. In order to derive these limits, we take into account the pure $Z'$-boson signal along with its interference with the SM background as suggested in Refs.~\cite{Accomando:2011eu,ACCOMANDO:2013ita,Accomando:2013dia,Accomando:2015rsa,Accomando:2013sfa,Accomando:2016mvz}. We closely follow the validated procedure given in Refs.~\cite{Accomando:2013sfa,Accomando:2015ava} where we have included the acceptance times efficiency factors for the electron and muon DY channels quoted by the CMS analysis \cite{Khachatryan:2016zqb}. We have combined the two channels and used  Poisson statistics to extract the 95\% CL bounds in the $\tilde g, g'_1$ plane for different $Z'$-boson masses.
See also \cite{Alves:2015mua,Klasen:2016qux} for related analyses.
Taking the allowed values of $g'_1$ and $M_{Z'}$, we can then compute $x$. For three $x$ benchmark values and for $\alpha$ = 0.3, in Fig.~\ref{Fig.SigmaBr}(a), we plot the $\sigma \times \textrm{Branching~Ratio~(BR)}$ for the process $pp \rightarrow H_1 \rightarrow \nu_h \nu_h$ at the 13 TeV LHC as a function of the heavy neutrino mass. The chosen value of the scalar mixing angle complies with the exclusion bounds from LEP, Tevatron and LHC searches and is compatible with the measured signals of the discovered SM-like Higgs state (hereafter, taken to have a mass of 125 GeV) in most of the interval $150 \, \textrm{GeV} \leq m_{H_2 }\leq 500\, \textrm{GeV}$ \cite{Accomando:2016sge}. The previous constraints are enforced using the \texttt{HiggsBounds} \cite{arXiv:0811.4169,arXiv:1102.1898,arXiv:1301.2345,arXiv:1311.0055,arXiv:1507.06706} and \texttt{HiggsSignals} \cite{Bechtle:2013xfa} tools. A high heavy neutrinos production rate can be obtained for low $x$-values. These values correspond to large $g'_1$ couplings, which are more likely to be allowed for higher $Z'$ masses. As an example, for $x$ = 4 (corresponding to an allowed point defined by $M_{Z'}$ = 4 TeV and $g'_1$ = 0.5), we get a  cross section of 307 fb. The cross section goes down by decreasing the scalar mixing angle. 

\noindent
The BR of the light SM-like Higgs into heavy neutrinos decreases with decreasing mixing angle $\alpha$, while its gluon-induced production rate increases. The net effect is that the cross section $\sigma (pp\rightarrow H_1\rightarrow \nu_h\nu_h )$ diminishes with $\alpha$. 
For the same parameter point above ($M_{Z'}$ = 4 TeV and $g'_1$ = 0.5), lowering the value of $\alpha$ from 0.3 to 0.1 reduces the cross section by roughly a factor of ten. For 
$\alpha$ = 0.1, we get in fact about 35 fb.

\begin{figure}
\centering
\subfigure[]{\includegraphics[scale=0.54]{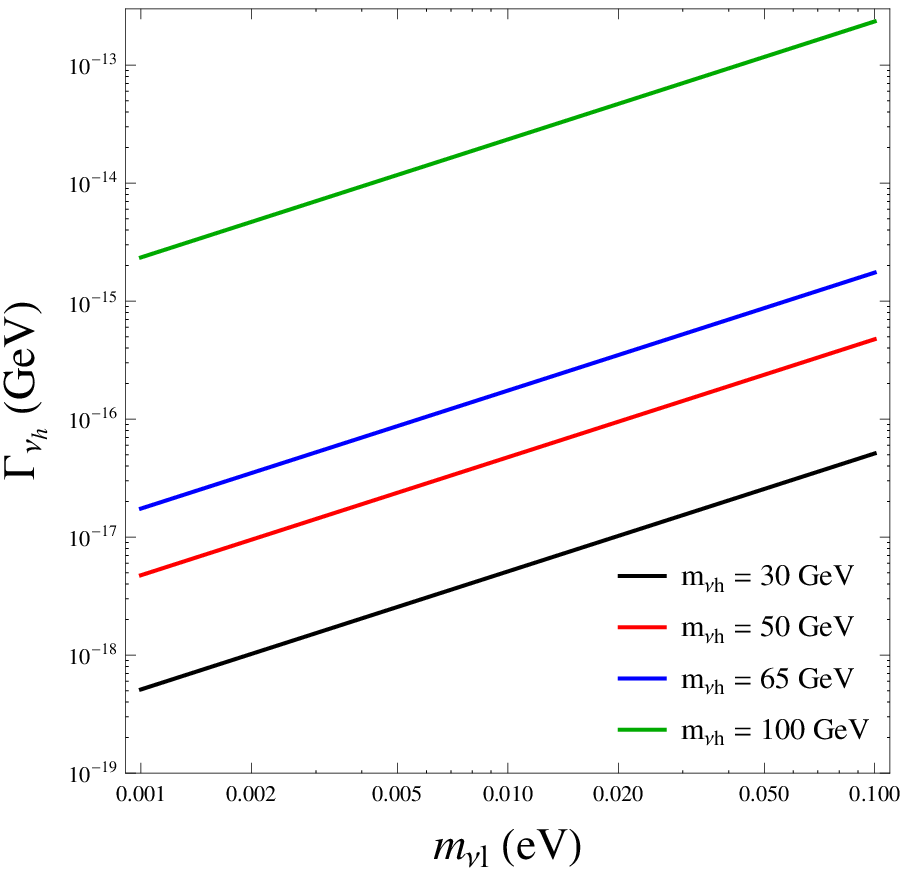}}
\subfigure[]{\includegraphics[scale=0.535]{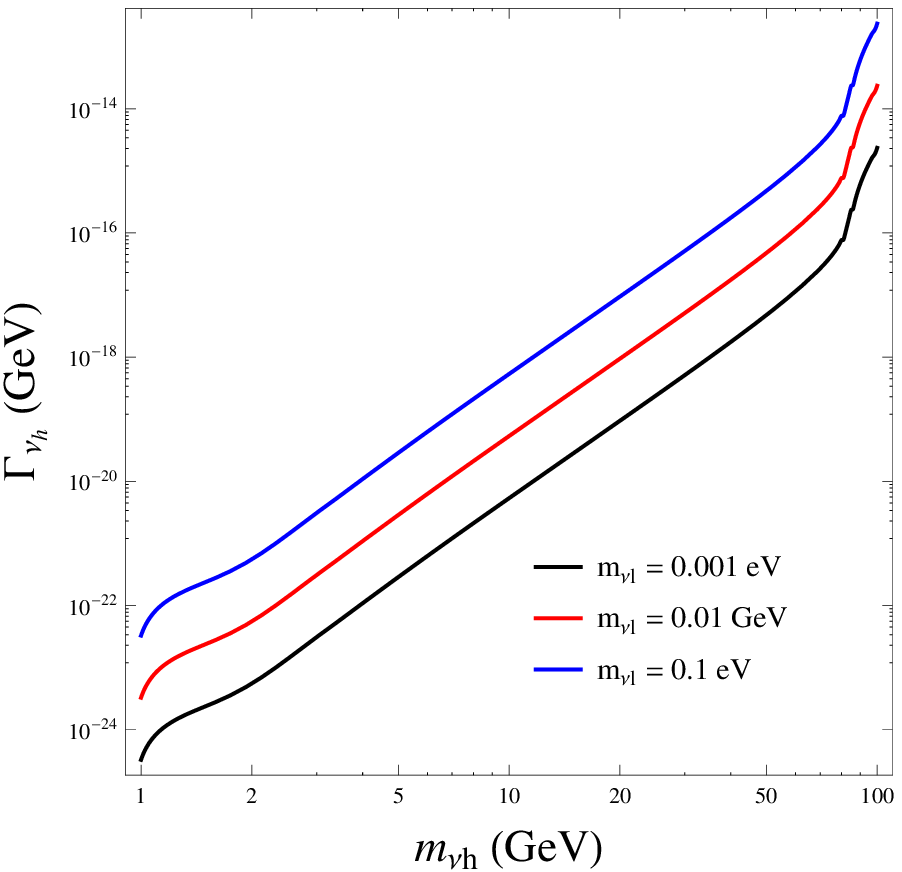}}
\subfigure[]{\includegraphics[scale=0.385]{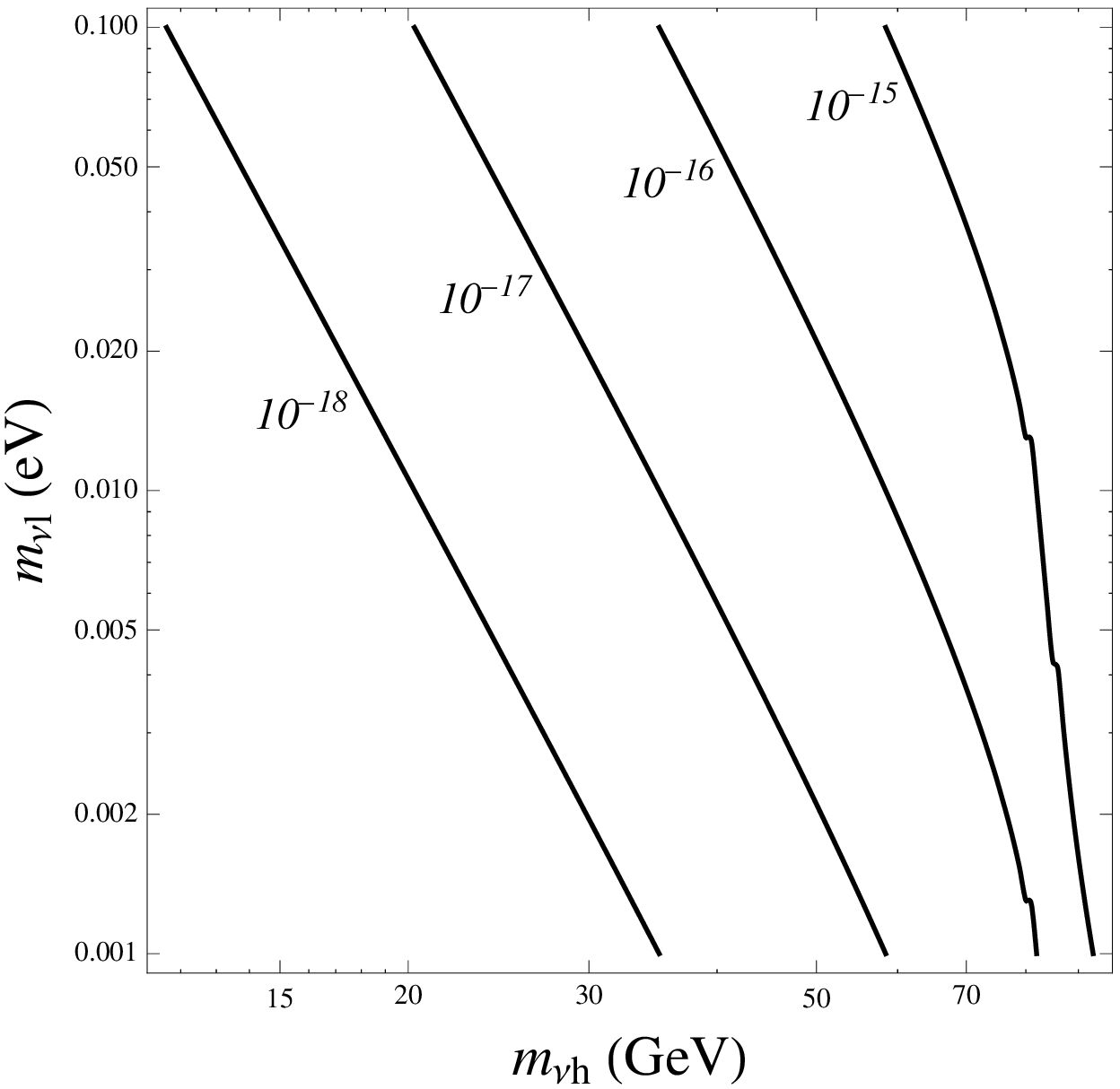}}
\caption{Total decay width of the heavy neutrino at rest as a function of (a) the light neutrino mass,  $m_{\nu_l}$, and of (b) the heavy neutrino mass, $m_{\nu_h}$. (c) Contour lines of the heavy neutrino decay width at rest (in GeV) in the plane $(m_{\nu_h}, m_{\nu_l})$. \label{Fig.Gamma}}
\end{figure}

\begin{figure}
\centering
\subfigure[]{\includegraphics[scale=0.54]{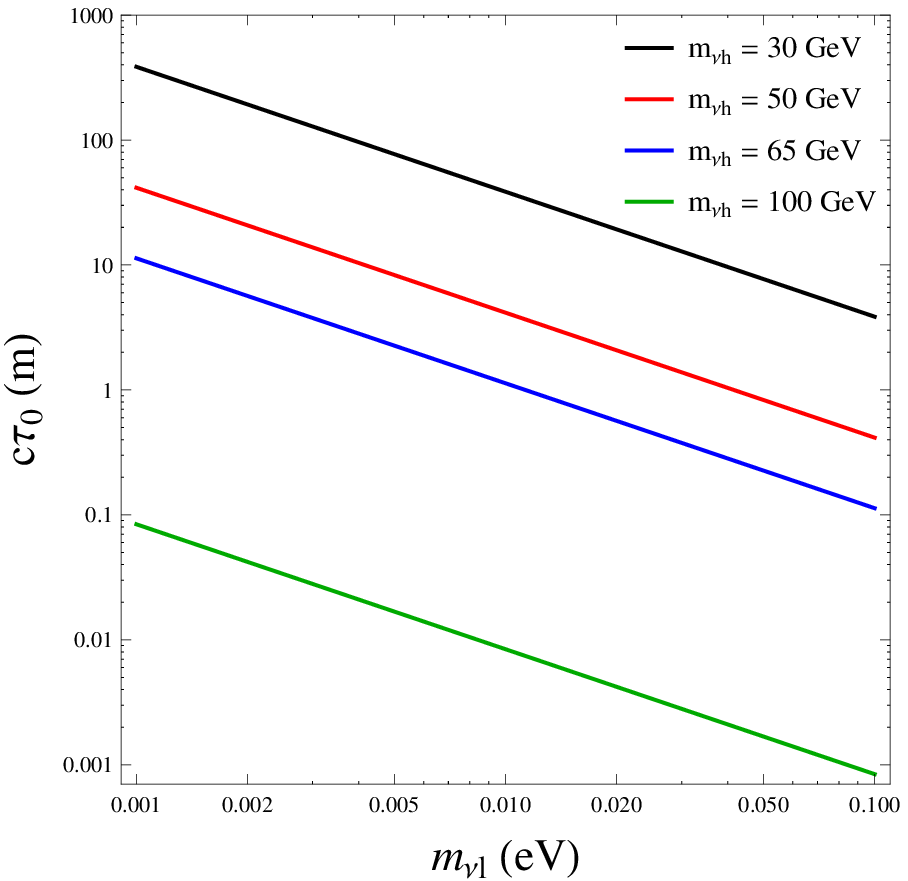}}
\subfigure[]{\includegraphics[scale=0.54]{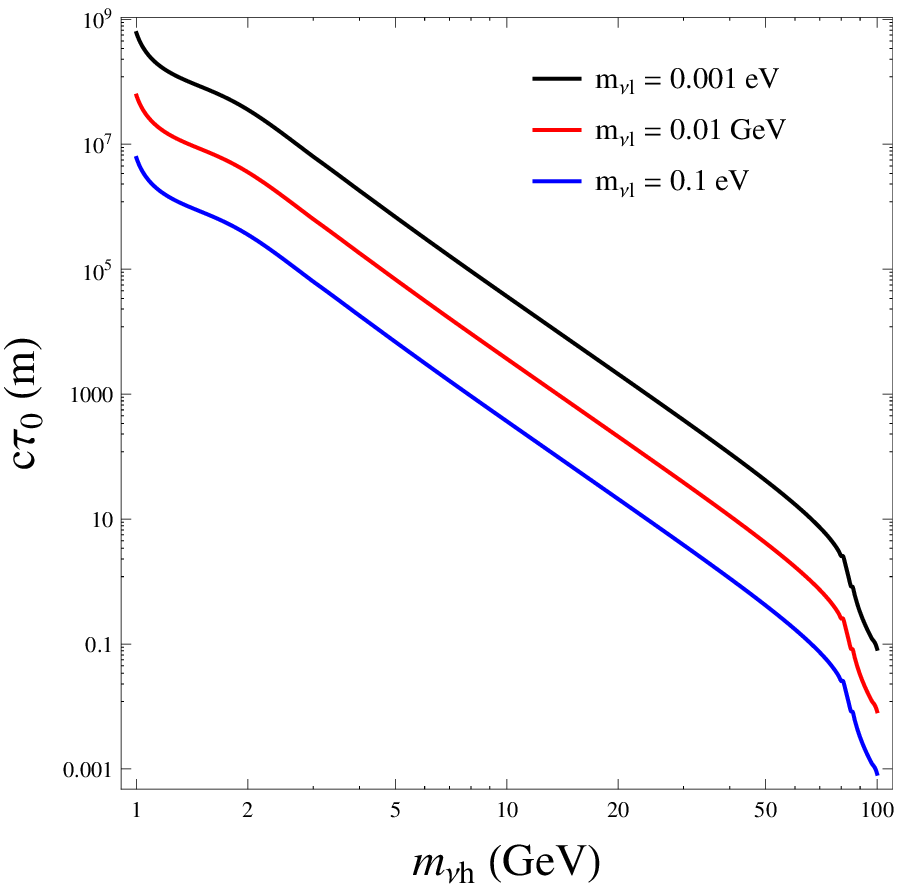}}
\subfigure[]{\includegraphics[scale=0.39]{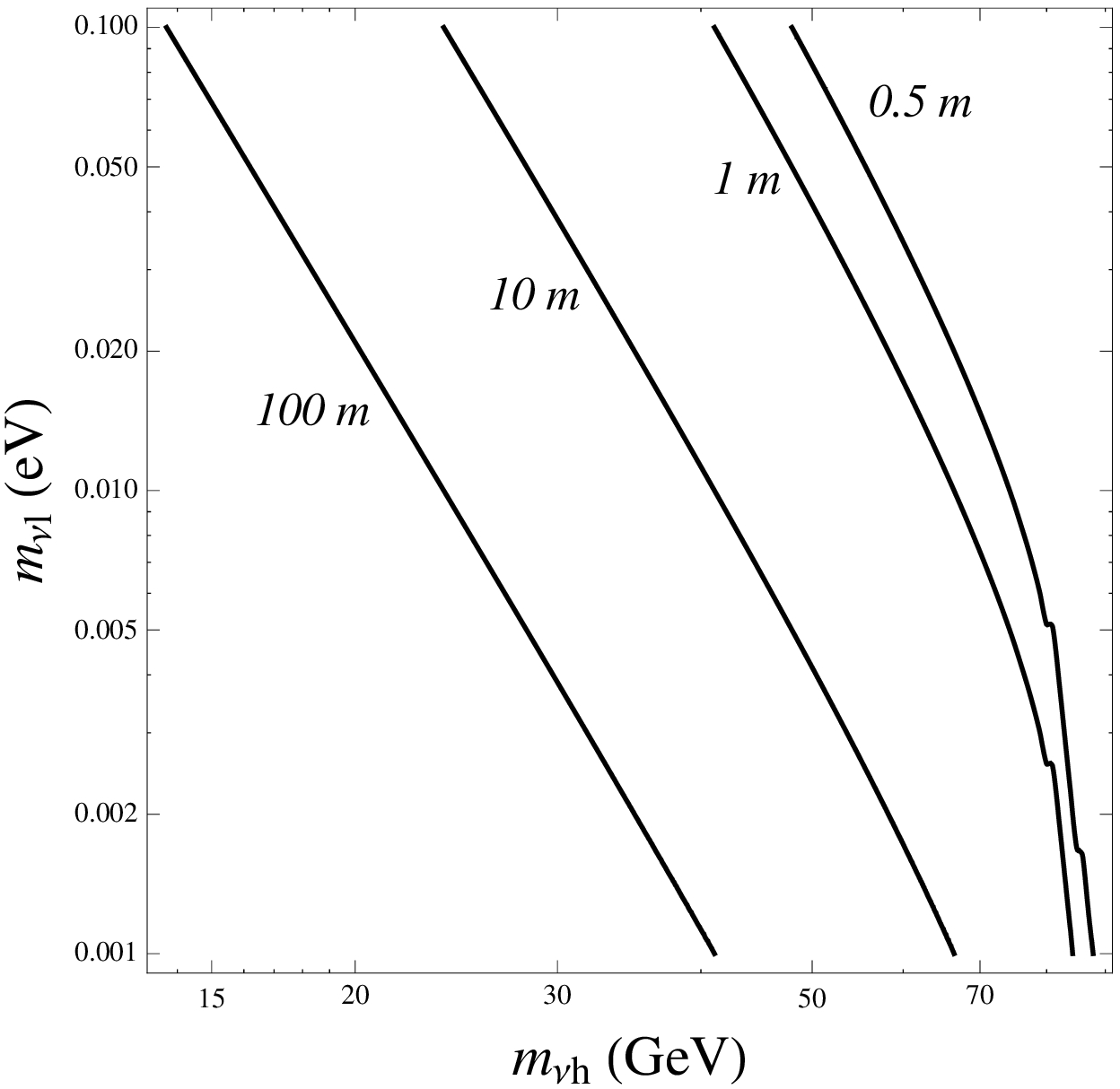}}
\caption{Proper decay length of the heavy neutrino as a function of (a) the light neutrino mass, $m_{\nu_l}$, and of (b) the heavy neutrino mass, $m_{\nu_h}$. (c) Contour lines of the heavy neutrino proper decay length in the plane $(m_{\nu_h}, m_{\nu_l})$. 
\label{Fig.DecayLength}}
\end{figure}

\noindent
In the mass region $m_{\nu_h} \le m_{H_1}/2$, the heavy neutrinos undergo the following decay processes: $\nu_h \rightarrow l^\pm \, W^{\mp *}$ and $\nu_h \rightarrow \nu_l \, Z^*$ with off-shell gauge bosons. In principle, $\nu_h \rightarrow \nu_l \, H_{1,2}^*$ and $\nu_h \rightarrow \nu_l \, Z^{'*}$ could also be activated but their BRs are extremely small in the kinematic region considered here. In Fig.~\ref{Fig.SigmaBr}(b) we show the BR of the heavy neutrinos for the four available modes $l^+l^- \nu_l$ (black), $qql$ (blue), $qq \nu_l$ (red) and $\nu_l\nu_l\nu_l$ (green), where $l=e,\mu,\tau$ and $q=u,d,s,c,b$. The first channel, which accounts for the 23\% of the decay modes, is mediated by both the $W^\pm$ and the $Z$ boson, the second reaches 50\% and is produced solely by $W^\pm$ exchange while the last two are determined by the $Z$ alone. As we have already discussed, the couplings of the heavy neutrinos to the gauge bosons are proportional to $V_{\alpha i}$ and, therefore, are extremely small leading to very long decay lengths $c \tau_0$, where $\tau_0$ is the heavy neutrino lifetime in the rest frame. These long lived particles may generate DVs inside the detector. We show in Figs.~\ref{Fig.Gamma} and \ref{Fig.DecayLength} the total decay width of the heavy neutrino in the rest frame and its proper decay length as a function of the light (a plots) and the heavy (b plots) neutrino mass, respectively. As the total decay width is proportional to $V_{\alpha i}^2 m_{\nu_h}^5 = m_{\nu_l}^2 m_{\nu_h}^3$, it decreases by lowering either of the two masses as illustrated in Fig.~\ref{Fig.Gamma}(c). The proper lifetime is the inverse of the decay width at rest. Thus, light enough RH neutrinos can be quite long lived. In Fig.~\ref{Fig.DecayLength}(c), we show the proper decay length $c\tau_0$ in the plane ($m_{\nu_h}, m_{\nu_l}$). In the next section, we see how this characteristic of heavy neutrinos will impact on their detection at the LHC. 

\noindent
Relaxing the degenerate mass hypothesis in the neutrino sector and taking into account the complete mixing matrix would make the analysis much more involved but the methodology would remain the same. The width and BRs of the heavy neutrinos have been computed with \texttt{CalcHEP} \cite{Belyaev:2012qa} using the $U(1)'$ model file \cite{Basso:2010jm, Basso:2011na} accessible on the High Energy Physics Model Data-Base (HEPMDB) \cite{hepmdb}.

\noindent
In addition, as $c \tau_0$ scales with $|V_{\alpha i}|^{-2} = m_{\nu_h}/m_{\nu_l}$, a simultaneous measurement of the decay length and of the mass of the heavy neutrinos could, in principle, provide insights on the elements of the mixing matrix and on the scale of the light neutrino masses, as previously mentioned. 


\subsection{Heavy neutrino decay probability and detector geometry}

The probability of the heavy RH neutrinos decaying in the detector depends on their  kinematic configuration and on the region of the detector in which the decay may occur. 
For definiteness, we focus our analysis on the CMS detector but the same reasonings apply unchanged to the case of the ATLAS experiment. 

The experimental efficiency for reconstructing a displaced event depends on the region of the detector where it decays. We take this into account by considering two regions. A region is chosen to select long lived heavy neutrinos that decay in the tracker beyond where tracks can be reconstructed and before the muon chambers to provide the possibility of observing muon hits (Region 1). A second region is chosen to select long lived heavy neutrinos that decay within the inner tracker (Region 2). Both regions can be approximated by a hollow cylinder. Region 1 is characterised by internal and external radii approximately given by $R_\textrm{min} = 0.5$ m and $R_\textrm{max} = 5$ m, plus longitudinal length, along the $z$ axis, given by $|z| < 8$ m. The requirement $r > R_\textrm{min}$ ensures that the muons are generated in a region in which the inner tracker track reconstruction efficiency is zero, while $r < R_\textrm{max}$ also guarantees that the heavy neutrinos decay in a region prior to the muon chambers. Additionally, Region 2 has $R_\textrm{min} = 0.1$ m, $R_\textrm{max} = 0.5$ m and  $|z| < 1.4$ m. The inner radius corresponds to a distance from the beam line which we define to be the lower limit, in the transverse plane, of the heavy neutrino decay vertex in order to safely neglect any source of SM background from the proton-proton collisions. 

\begin{figure}
\centering
\subfigure[]{\includegraphics[scale=0.35]{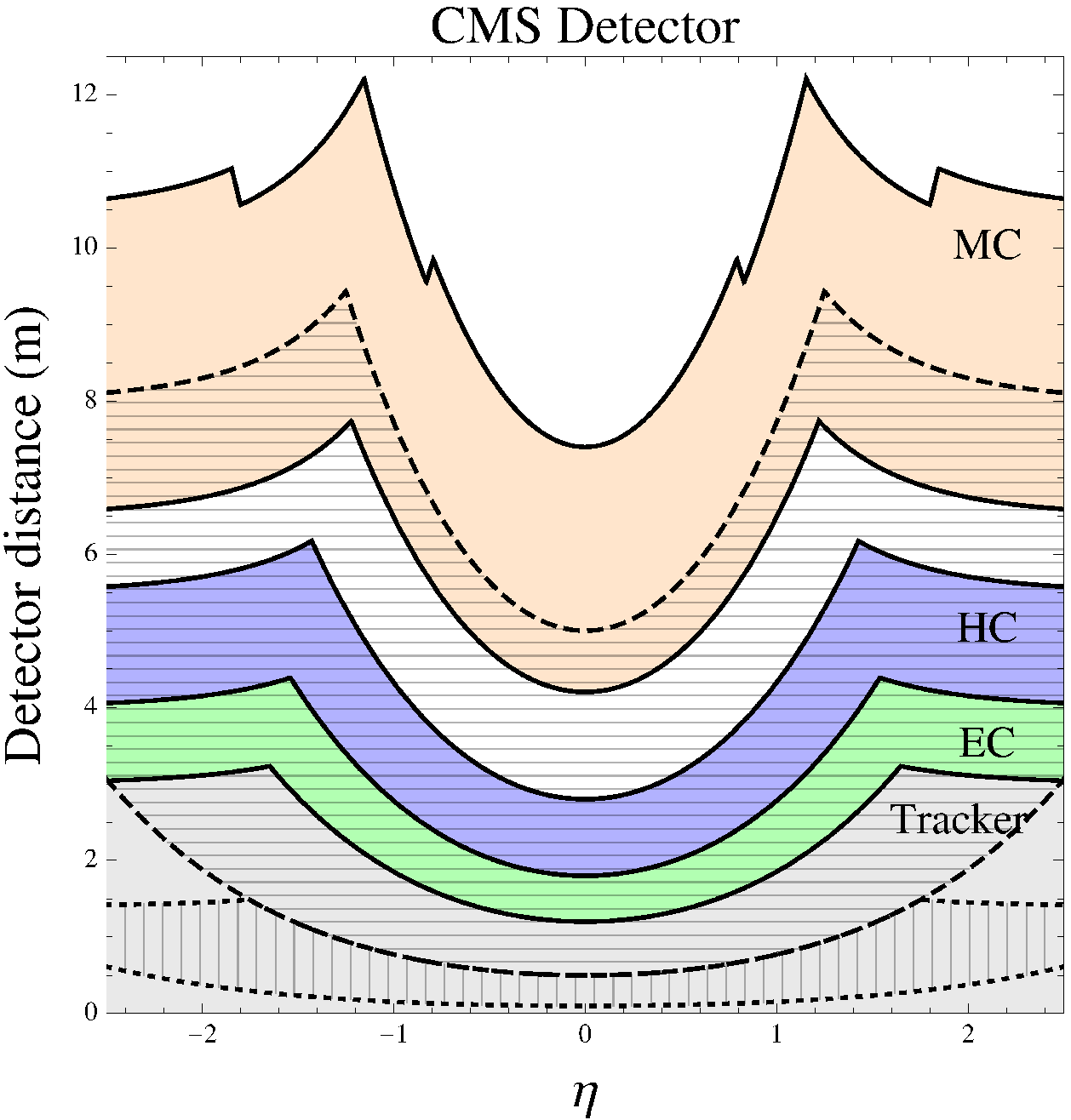}} \quad
\subfigure[]{\includegraphics[scale=0.49]{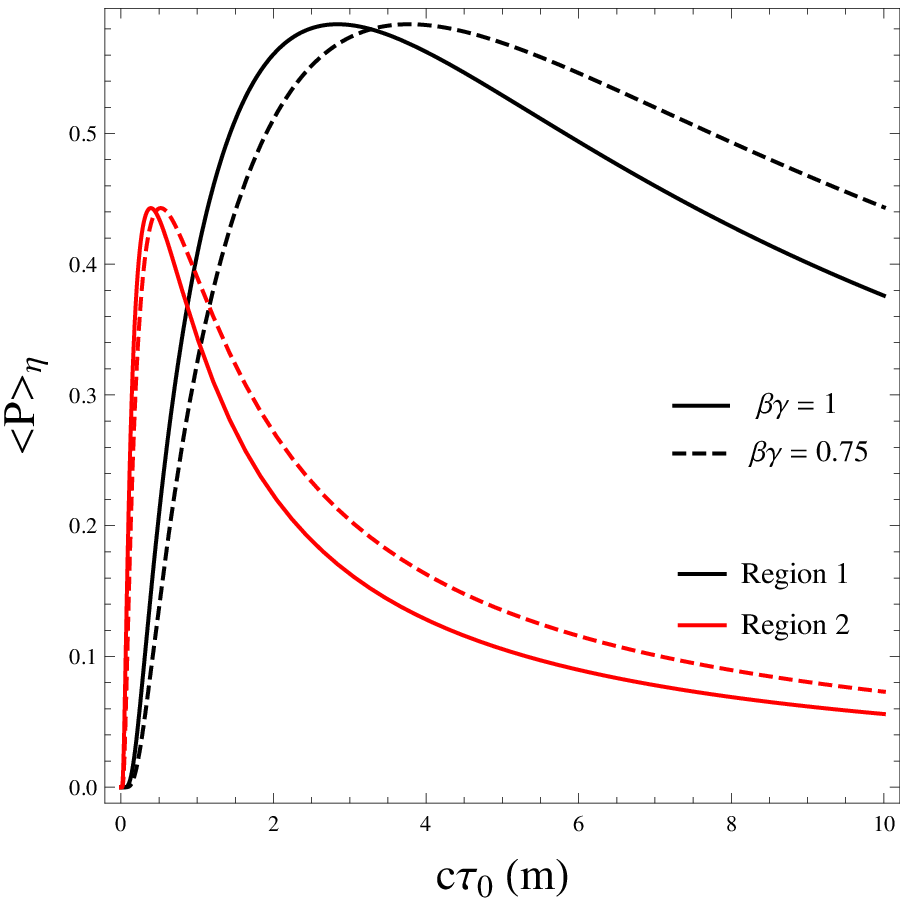}} \quad
\subfigure[]{\includegraphics[scale=0.49]{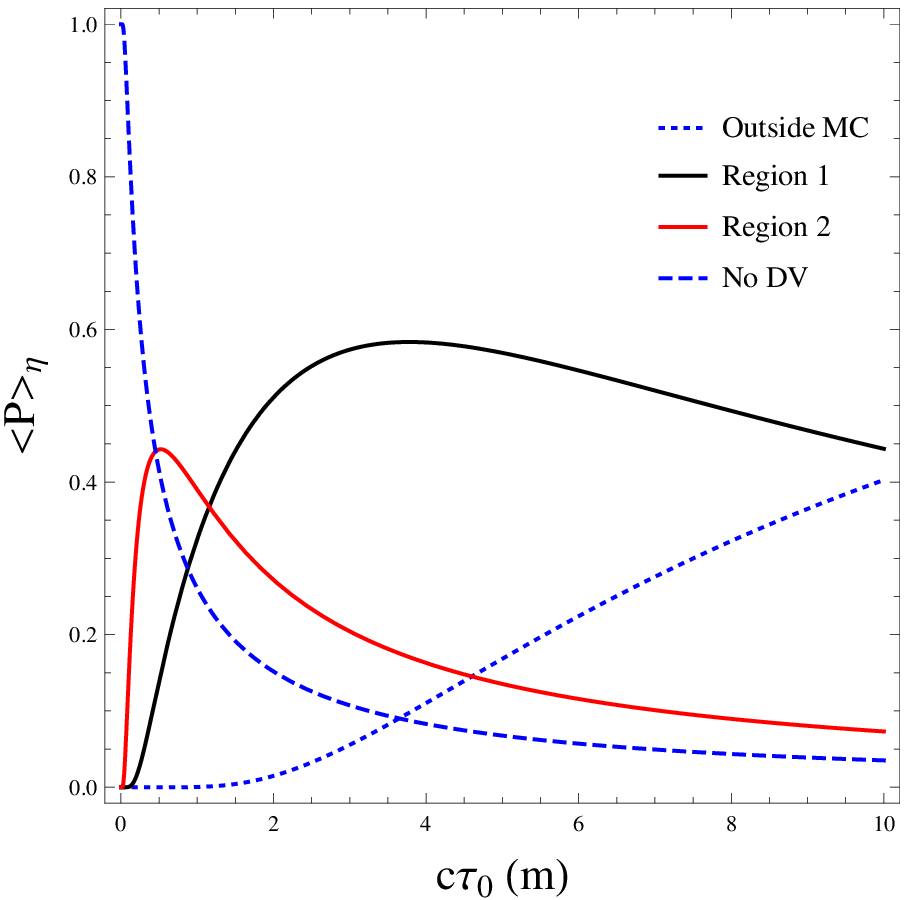}}
\caption{(a) Geometry of the CMS detector as a function of the pseudo-rapidity $\eta$. From bottom to top, one has the inner tracker (grey area), the EM calorimeter (green area), the hadronic calorimeter (blue area) and the muon chamber (orange area). The horizontal and vertical hatched areas correspond to Region 1 and 2, respectively  (see text). (b) Averaged decay probability of the heavy neutrino as a function of its proper decay length $c \tau_0$ for two values of the $\beta\gamma$ relativistic factor and for Region 1 and 2 of the detector, as explained in the text. (c) Averaged decay probability of the heavy neutrino for $\beta \gamma = 0.75$ in four different regions of the detector. \label{Fig.CMSstructure}}
\end{figure}

In Fig.~\ref{Fig.CMSstructure}(a) we depict an approximate description of the CMS detector \cite{Chatrchyan:2013sba} whose parts are described in terms of the distance covered by a possible heavy RH neutrino before decaying. Such a distance is a function of the neutrinos pseudo-rapidity, $\eta$ (or the associated scattering angle). In particular, we show the tracker (grey region), the EM calorimeter (green region), the hadronic calorimeter (blue region) and the muon chamber (orange region). The tracker has been described as a cylinder whose central axis coincides with the beam line, while the EM and hadronic calorimeters as well as  the muon chamber can be approximated as concentric cylinders. The segmented line defining the outer region of the muon chamber reflects the presence of the endcaps. The vertical and horizontal hatched areas correspond to Region 2 and 1, respectively, namely, the regions in which the heavy neutrinos decay is optimised for lepton and jet identification in the inner tracker and for muon detection in the muon chamber.  

The probability for the heavy neutrino decaying in the annulus defined by the radial distances $d_1(\eta)$ and $d_2(\eta)$ at the pseudo-rapidity $\eta$ is
\bea
P = \int_{d_1(\eta)}^{d_2(\eta)} d x \frac{1}{c \tau} \exp\left(- \frac{x}{c \tau}\right) \,,
\label{probability}
\eea
where $c \tau = \beta \gamma c \tau_0$ is the decay length in the lab frame with $\beta \gamma$ the corresponding relativistic factor. In order to understand the behaviour of the probability of heavy neutrino decaying as a function of the proper decay length $c \tau_0$ we consider the average probability $\langle P \rangle_\eta$ over an isotropic angular distribution of the heavy neutrinos. The relativistic factor $\beta \gamma$ can be estimated assuming that the Higgs is produced at rest in which case we obtain $\beta \gamma \simeq 0.75$ for $m_{\nu_h} = 50$ GeV. The results are shown in Fig.~\ref{Fig.CMSstructure}(b) for the two different decay volumes described above. The solid line with $\beta\gamma = 1$ represents the exact dependence of $P$ on the decay length $c \tau$ in the laboratory frame. Fig.~\ref{Fig.CMSstructure}(c) shows the averaged decay probabilities computed for $\beta \gamma \simeq 0.75$ outside Region 1 (dotted blue), inside Region 1 (black), inside Region 2 (red), and in the inner region of the tracker (dashed blue) where the vertices and tracks are not displaced. Notice that, for $c \tau_0 = 2$, the probability of decay inside the muon chambers and the inner detector are, approximately, 50\% and 25\%, respectively, with only a small fraction escaping the detector. For smaller decay lengths the results may be different. As an example, for $c \tau_0 = 0.5$, the heavy neutrinos predominantly decay in the tracker providing displaced and non-displaced signals with almost equal probabilities. The latter signature could be addressed, in principle, with standard, non-displaced, techniques at the cost of increased SM backgrounds. However, this is impractical due to the small heavy neutrino masses ($m_{\nu_h} < m_{H_1}/2$) and  poor efficiency of the lepton identification (see discussion below on the lepton $p_T$) which prevents discrimination of the signal event from the SM background. When the RH neutrino decays inside the inner tracker, we thus rely on the DVs whose signature has a negligibly small background contribution. For much larger values of the proper lifetime ($c\tau_0$ > 10 m), the RH neutrinos have a sizeable probability of decaying outside the detector. In this event, the heavy neutrino pair production would appear as a missing energy signature and techniques appropriate for the invisible Higgs decay should be applied (see for example Ref.~\cite{Belyaev:2015ldo} for details on such a technique). 
\section{Event simulation}
\label{sec:eventsimulation}
In this section we present a search for long-lived heavy neutrinos decaying into final states that include charged leptons (electrons and muons). The corresponding DVs can be reconstructed if the heavy neutrinos decay within the volume of the CMS inner tracker. However, here we also want to exploit the possibility to identify DVs beyond the region where they can be reconstructed in the tracker and prior to the muon chambers of the CMS detector (in this case only muons are taken into account).

In order to highlight the sensitivity of DVs due to the heavy neutrino decays produced by the 125 GeV Higgs, we present the details of a parton-level Monte Carlo (MC) analysis at the LHC with a Centre-of Mass  (CM) energy of $\sqrt{S} = 13$ TeV and luminosity $\mathcal L = 100$ fb$^{-1}$. 
For our simulation we select four Benchmark Points (BPs) characterised by different heavy neutrino masses and different proper decay lengths. These two quantities unambiguously fix the light neutrino mass. The other relevant parameters, $M_{Z'}, g'_1, \alpha$, are as follows: $M_{Z'} = 5$ TeV, $g' _1= 0.65$ and $\alpha = 0.3$. The proper decay length of the heavy neutrinos is chosen to optimise their observability in different regions of the CMS detector, in particular, BP1 and BP2 are characterised by heavy neutrinos mostly decaying in the muon chamber, while BP3 and BP4 provide long-lived particles that are better identified in the inner tracker. The different BPs are defined in Tab.~\ref{Tab:H1BPs}.
\begin{table}
\centering
\begin{tabular}{|c||c|c|c||c|}
\hline 
 	&  $m_{\nu_h} \, \textrm{(GeV)}$ & $m_{\nu_l} \, \textrm{(eV)}$ & $c \tau_0 \, \textrm{(m)}$ & $\sigma_{\nu_h\nu_h} \, \textrm{(fb)}$ \\ \hline
   BP1 &  40 & 0.075 & 1.5 & 332.3 \\  
   BP2 &  50 & 0.02 & 2.0 & 248.3 \\  \hline
   BP3 &  45 & 0.065 & 1.0 & 310.2 \\ 
   BP4 &  50 & 0.082 & 0.5 & 248.3 \\  \hline
\end{tabular}
\caption{Definition of the BPs leading to DVs from the heavy neutrinos produced by the SM-like Higgs decay. From left to right, the columns display the values of heavy RH neutrino mass, light neutrino mass, proper decay length of the heavy neutrinos and production rate of heavy neutrino pairs coming from the decay of the SM-like Higgs produced via gluon fusion. \label{Tab:H1BPs}}
\end{table}
As leptons can originate from the three-body decay of soft heavy neutrinos, $m_{\nu_h} < m_{H_1}/2 \simeq 62.5$ GeV, a cut on their transverse momentum is found to have the greatest effect among all the kinematic acceptance requirements. We show in  Tab.~\ref{tab.ptleptoncuts} the efficiencies of the $p_T$ cut on leptons for different thresholds in the BP1 scenario. In particular, we report the combined cuts for different thresholds $p_T^{(1)}$ on the two leading leptons and different thresholds $p_T^{(2)}$ on the third sub-leading lepton, showing how the efficiency drops to a few percent for $p_T^{(1)} \simeq 26$ GeV, which is a trigger requirement at CMS for both the muons reconstructed in the muon chamber and the leptons identified in the inner tracker. These results suggest that a dedicate trigger with a lower lepton $p_T$ threshold would be particularly useful for the analysis of rather soft and long-lived particles. 

We consider the process $pp \rightarrow H_1 \rightarrow \nu_h \nu_h$ with the two heavy neutrinos decaying into two to four muons for the analysis of DVs in the muon chamber and into two to four leptons (electrons and muons) for the corresponding analysis in the inner tracker. We do not reconstruct jets in the event. The individual decay chains of the heavy neutrino are summarised as follows:
\begin{itemize}
\item $\nu_h\rightarrow l^\mp W^\pm\rightarrow l^\mp {l'}^\pm\nu_{l'}$
\item $\nu_h\rightarrow l^\mp W^\pm\rightarrow l^\mp q\bar{q'}$
\item $\nu_h\rightarrow \nu_{l'} Z\rightarrow \nu_{l'} l^+ l^-$
\item $\nu_h\rightarrow \nu_{l'} Z\rightarrow \nu_{l'} q\bar q$
\item $\nu_h\rightarrow \nu_{l'} Z\rightarrow \nu_{l'} \nu_l \nu_l$
\end{itemize}
where $l = e, \mu, \tau$ and $q$ can be one of five flavours ($q = d, u, c, s, b$). The $\tau$ lepton can decay into $e, \mu$ and hadrons. When considering the decay of the heavy neutrino pair, one can have signatures with two same-sign leptons of same or different flavour, signatures with two opposite-sign leptons of same or different flavour and signatures with more than two leptons.  

\noindent
For each event, we evaluate the length $c\tau$ in the laboratory for the two RH neutrinos. This length depends on the heavy neutrino speed via the relativistic factor $\beta\gamma$. We then randomly sample the distance $L$ travelled by each of the two heavy neutrinos from the exponential distribution $\exp(-x/c\tau)$. Using the simulated momentum of the two heavy neutrinos (or the scattering angles), we can then determine the position of the two DVs. Standard acceptance requirements are imposed on the transverse and longitudinal decay lengths of the heavy neutrinos $L_{xy}$ and $L_z$, respectively. As stated in \cite{CMS:2015pca}, in order to identify muons in the muon chambers, each of the reconstructed muon tracks must satisfy $|L_z| < 8$ m, $L_{xy} < 5$ m and $L_{xy}/\sigma_{L_{xy}} > 12$. The resolution $\sigma_{L_{xy}}$ on the transverse decay length is approximately 3 cm. Thus the geometrical constraints are $0.36 \, \textrm{m} < L_{xy} < 5$ m, globally. The lower and upper bounds ensure that the muons are generated in a region where they are not reconstructed by the tracker and can be identified by the muon chambers. Furthermore, the identification of leptons in the inner tracker demands $0.1 \, \textrm{m} < L_{xy} < 0.5 \, \textrm{m}$ and $|L_z| < 1.4$ m \cite{CMS:2014hka,Khachatryan:2014mea}. The upper bound on $L_{xy}$ is determined by the efficiency of the tracker while the lower bound provides DVs in a region where the contamination from the SM background is negligible.

\noindent 
In addition, two extra cuts on the impact parameter, $d_0$, and  the angular separation between muon tracks, $\theta_{\mu\mu}$, are used to completely suppress any source of SM background. To this end, we closely follow Ref.~\cite{CMS:2015pca} and we first implement generic detector acceptance requirements to identify the leptons (electrons and muons). In particular, we impose $|\eta_l|$ < 2, $\Delta R_l > 0.2$, $p_T^l > 26$ GeV for the two leading leptons and $p_T^l > 5$ GeV for any sub-leading leptons. 
Notice that, differently from Ref.~\cite{CMS:2014hka} where a Run I-designed trigger has been adopted, we use trigger thresholds potentially available in Run II where both leading electrons and muons can be required to have $p_T > 26$ GeV.
Once the lepton tracks are reconstructed, we then implement the two additional cuts on $\theta_{\mu\mu}$ and $d_0$.
A significant source of background may arise from cosmic ray muons which may be misidentified as back-to-back muons. Such events are removed by requiring that the opening angle between two reconstructed muon tracks, $\theta_{\mu\mu}$, satisfies the constraint $\cos \theta_{\mu\mu} > - 0.75$.  
The impact parameter $d_0$ in the transverse direction of each of the reconstructed tracks of electrons and muons (identified by the acceptance requirements described above) is given by the expression
\bea
|d_0|=|x \, p_y - y \, p_x|/p_T,
\eea
where $p_{x,y}$ are the transverse components of the momentum and $p_T$ is the transverse momentum of the leptonic track. In the above formula, the variables $x$ and $y$ correspond to the position where the heavy RH neutrino decays. They are determined by the projection on the $x, y$ transverse plane of the length $L$ covered by the neutrino before decaying.
 
\begin{figure}
\centering
\includegraphics[scale=0.37]{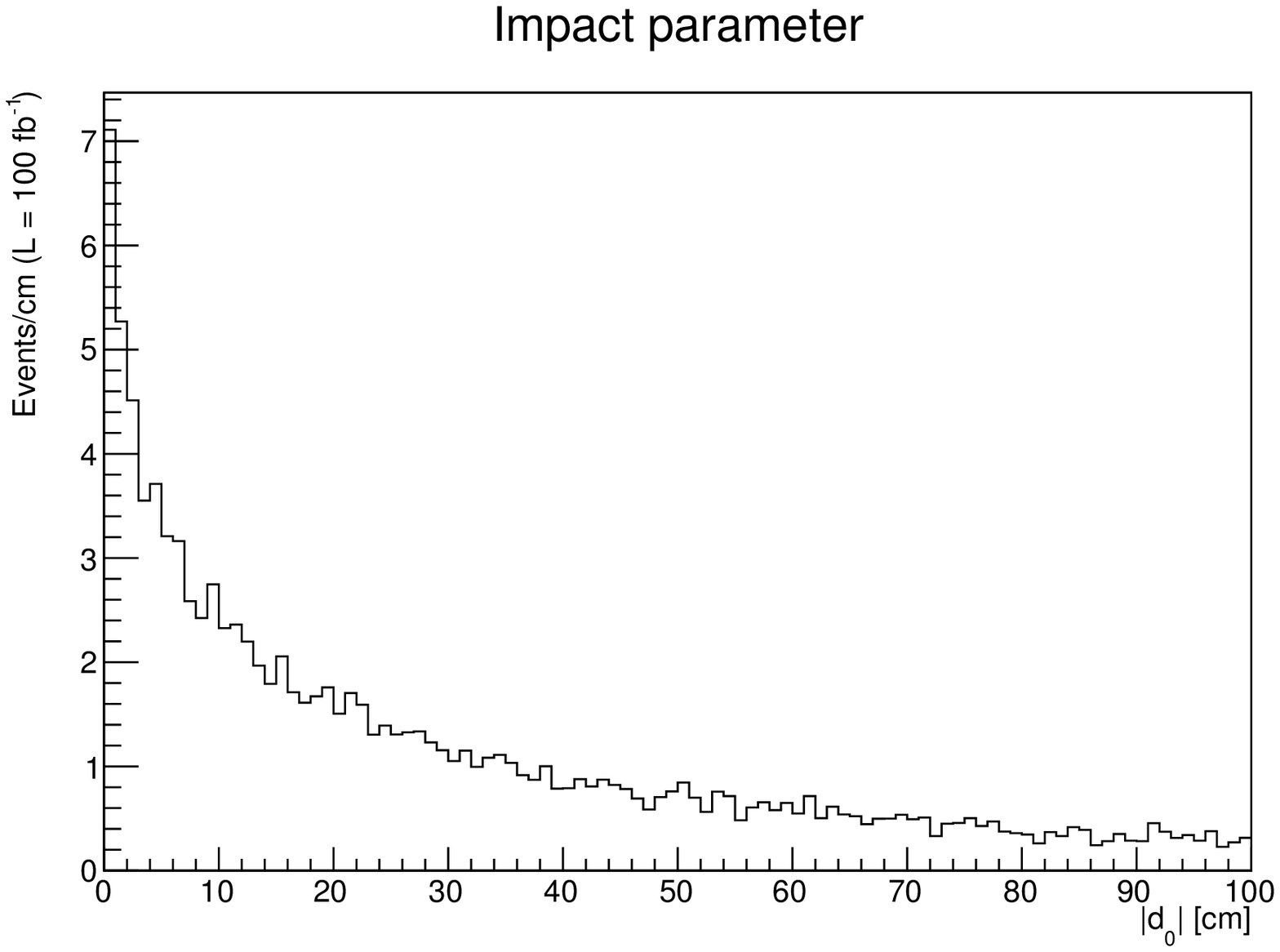}
\includegraphics[scale=0.37]{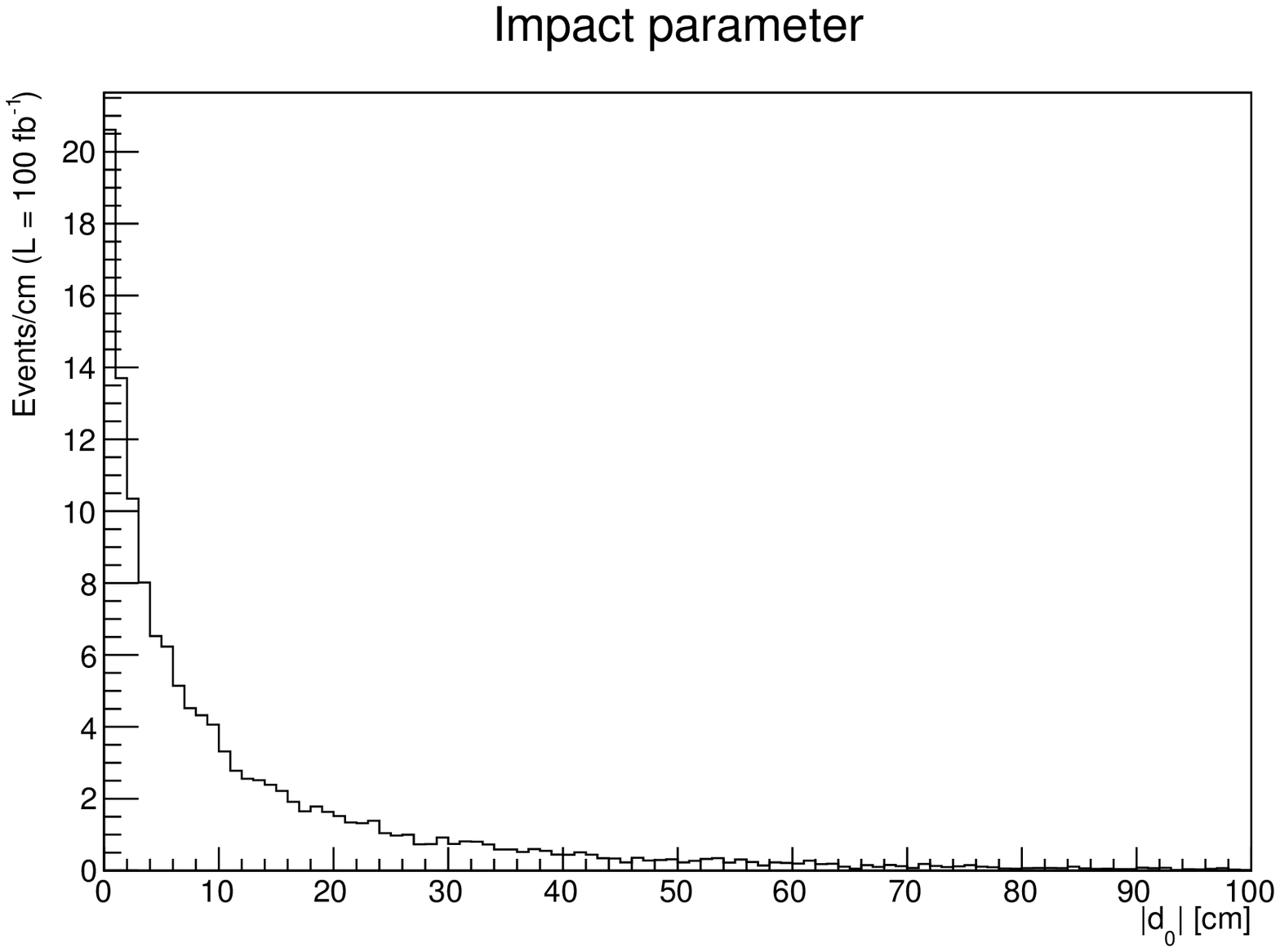}
\caption{Impact parameter distribution for (a) BP2 and (b) BP4 of Tab. \ref{Tab:H1BPs}. The heavy neutrino proper decay lengths are, respectively, 2 m and 0.5 m.\label{Fig:impactparam}}
\end{figure}

In Fig.~\ref{Fig:impactparam}, we show the $d_0$ distribution for BP2 and BP4  in which identification cuts on the leptons are applied. As expected, BP4 provides a tighter distribution characterised by a shorter heavy neutrino lifetime. According to \cite{CMS:2015pca,CMS:2014hka}, the SM background can be significantly reduced by applying cuts on the impact parameter significance $|d_0|/\sigma_{d}$ where the resolution $\sigma_d$ is approximately given by 2 cm and 20 $\mu$m in the muon chambers and the tracker, respectively. We thus impose $|d_0|/\sigma_{d} > 4$ for the muons in the muon chamber and $|d_0|/\sigma_{d} > 12$ for the leptons in the inner tracker.

\noindent
Finally, the reconstruction efficiency $\epsilon$ of the lepton momenta is taken into account. A reasonable choice of $\epsilon$ for a single lepton, electron or muon, is $\epsilon = 90 \%$. The effect of these cuts and the results of the analysis are discussed in the following section and summarised in Tabs.~\ref{tab:MC_H1} and \ref{tab:IT_H1}. 

\subsection{Discussion of the results}
In this section, we present the results of the MC analysis of displaced leptons. We commence with the analysis of displaced muons reconstructed using only the muon chamber.
 
\subsubsection{Muon chamber}
It is instructive to classify the events in three distinct categories defined by the number of identified muons, from 2 to 4. In this analysis we use detector trigger requirements which tag two leptons. The cut flow for the different  BPs is depicted in Tab.~\ref{tab:MC_H1}.
Two muons of the $2\mu$ category appear in the muon detector as distinct tracks (they are generated from two different heavy neutrinos) which do not join in a single DV (the number of events with two muons coming from a single DV is less than one). This is due to the combined impact of the $p_T$ cut on the muons and to the small mass of the heavy neutrinos which reduce the probability of selecting two high-$P_T$ leptons originating from the same heavy neutrino. The $3\mu$ category, instead, is characterised by two muon tracks forming a single DV (one high-$P_T$ and one low $P_T$ muon track) and by a third separate high-$P_T$ track, while the $4\mu$ class provides two distinct DVs. Due to the kinematical features discussed above, a DV is always formed by one of the two most energetic muons ($p_T > 26$ GeV) and one with low $p_T$ ($5 \, \textrm{GeV} < p_T < 26 \, \textrm{GeV}$).  

The number of expected events for the four selected benchmark points of Tab.~\ref{Tab:H1BPs}, obtained after all cuts and efficiencies have been applied, is listed in Tab.~\ref{tab:MC_H1}. As expected,  BP4 leads to the smallest number of events among the different BPs. Indeed, even though its cross section and the heavy neutrino mass are the same as BP2, the short heavy neutrino decay length $c\tau_0 = 0.5$ m means the decay of the heavy neutrinos in the muon chambers is less likely and therefore reduces the sensitivity of this analysis to this particular benchmark point. The largest sample of events are collected within the BP1 scenario, where the big cross section and the large decay length enhance the number of muon tracks identified in the detector. Summing all the events in the three disjoint categories ($2\mu$, $3\mu$ and $4\mu$), the total expected number of events (after reconstruction and cuts) from heavy neutrinos is about $N_{evt}$ = 33 at $\mathcal L=100$ fb$^{-1}$ in the BP1 case. This signal has the benefit of having a negligibly small background contribution. The remaining two scenarios, BP2 and BP3, predict roughly 20 events each.

Concerning the possible source of backgrounds (apart from the cosmic muons) in the study of DVs in the muon chamber, the CMS analysis \cite{CMS:2015pca} takes into account muon pairs from DY which can be misidentified as displaced event due to detector resolution effects, production and decay of tau pairs giving rise to muons and $t\bar t$, $W^+W^-$, $ZZ$ and QCD multi-jets events.
In \cite{CMS:2015pca}, it is explicitly stated that all these backgrounds produce negligible contributions.

%
The signal events above could then be potentially detected at the LHC at the end of Run 2 data taking when the collected luminosity is expected to reach $\mathcal L=100$ fb$^{-1}$ at CMS.
\subsubsection{Inner tracker}

The same analysis can be performed exploiting the reconstructions of both electrons and muons in the inner tracker. As usual, we classify the events in three categories according to the number of leptons. The results are shown in Tab.~\ref{tab:IT_H1}. 

As explained above, the two leptons in the $2l$ category do not form a single DV but rather appear as separate tracks in the tracker. All the properties concerning the topologies of the different categories and the corresponding kinematical properties of the displaced leptons which have been discussed above for the analysis in the muon chamber apply equally here.
The potential sources of SM backgrounds listed in the previous section are also relevant in the analysis of DVs in the tracker. As shown in \cite{CMS:2014hka} and previously stated, cutting on the impact parameter successfully suppresses the SM background.

 After reconstruction and cuts, BP3 and BP4 are the best scenarios owing to the lower proper decay length $c \tau_0$ and, therefore, larger probability of the heavy neutrinos decaying inside the tracker volume (see Fig.~\ref{Fig.CMSstructure}(b)), compared to the other two benchmark points. Moreover, even though BP4 is characterised by a smaller cross section $\sigma_{\nu_h \nu_h}$ than BP3, the smaller decay length $c\tau_0$ allows for a larger number of heavy neutrinos to decay in the inner tracker, thus compensating and providing more reconstructed leptons than in the BP3 scenario. Tab.~\ref{tab:IT_H1} summarises the final number of DVs events expected in the inner tracker after reconstruction and cuts. As anticipated, 
the best scenario is BP4 with about 53 expected events, followed by BP3 with roughly 30 events. For the other two BPs less events are expected, providing a sample of 10 events each.

Being able to distinguish electrons from muons, in the inner tracker it is also possible to identify the flavour composition of the displaced tracks and vertices. We show the composition of the $2l$ and $3l$ events in Tab.~\ref{tab:IT_H1_2-3flavour} and of the $4l$ events in Tab.~\ref{tab:IT_H1_4flavour}. We use the notation $(f_1 f_2)$ to denote that the DV is made of two leptons of flavour $f_1$ and $f_2$. In the $(f_1 f_2)l$ case the lepton $l$ appears as a separate track.  

As clear from Tab.~\ref{tab:IT_H1_2-3flavour}, the fraction of events with two electrons is always slightly larger than the corresponding $\mu\mu$ case due to the presence of the cut used for the suppression of cosmic muons. The number of events in the mixed case $e \mu$ of the $2l$ category is suppressed with respect to the same-flavour leptons. This effect is due to our simplified setup with a diagonal Dirac Yukawa matrix $Y_\nu$ in which the two leptons originating from the first step of the two heavy neutrinos decay chains, namely $\nu_h \rightarrow  l^\pm \, W^{\mp *}$\footnote{The decay pattern $\nu_h \rightarrow  \nu_l \, Z^*$ undergoes a different counting but is suppressed with respect to the charged-current process.}, must share the same flavour, thus reducing the number of possible $e \mu$ pairs with respect to $ee$ or $\mu\mu$. Notice that this result is subject to the assumptions on the Dirac mass matrix. In contrast, the flavour composition of the DV of the $3l$ category is independent of the structure of $Y_\nu$. Interestingly, the majority of the events lay in the $(e \mu)l$ class which could represent a significant signature of DVs from heavy neutrinos.

\subsubsection{Comments on tri-leptons triggers}
\begin{figure}
\centering
\includegraphics[scale=0.6]{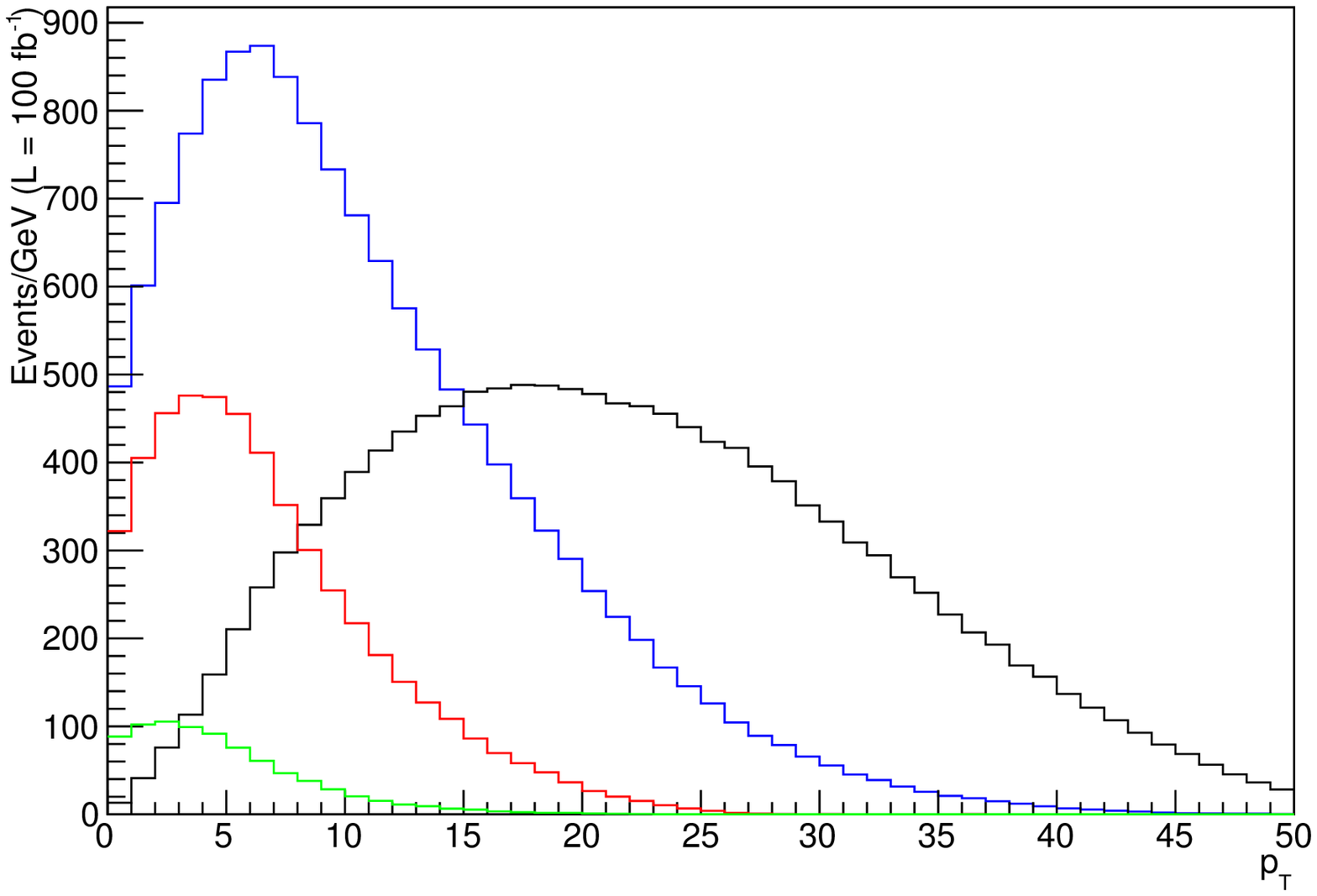}
\caption{Distributions of the leptonic transverse momentum. The colours black, blue, red and green correspond respectively to the momentum distributions of the four leptons ordered in decreasing $p_T$. \label{Fig:pt}}
\end{figure}

We show in Fig.~\ref{Fig:pt} the transverse momentum distributions of the four leptons where no cuts have been imposed. We take as reference the benchmark point BP1 of Tab.~\ref{Tab:H1BPs}. The different locations of the peaks suggest that an asymmetric choice for the $p_T$ thresholds of the four leptons could help to increase the efficiency of the selection. We comment on this possibility focusing on the implementation of an event identification for which at least three leptons are required. \\
We present in Tabs.~\ref{tab:IT_H1_3l1} and \ref{tab:IT_H1_3l2} the results of the DV analysis in which we have employed a tri-leptons trigger for long lived decays. This trigger differs from typical tri-lepton triggers in that it does not require any of the leptons to point back to the beamspot. In this case the thresholds on the lepton $p_T$ can be relaxed, with respect to the identification requirements discussed in the previous section, allowing for a larger number of events to be collected at the end of the selection procedure. Tri-lepton triggers have been extensively used in searches for supersymmetric particles but never employed in the study of displaced vertices. Being unable to refer to an existing literature, we examine the impact of two reasonable trigger requirements on the lepton $p_T$. The results in Tab.~\ref{tab:IT_H1_3l1} are derived requiring $p_T > 20$ GeV for the two most energetic leptons and $p_T > 10$ GeV for the third one, while Tab.~\ref{tab:IT_H1_3l2} is obtained considering $p_T > 20$ GeV for the most energetic lepton and $p_T > 15$ GeV for the other two. The fourth lepton, if present, is characterised by $p_T > 5$ GeV. All the other cuts remain unchanged. In Tabs.~\ref{tab:IT_H1_3l1} and \ref{tab:IT_H1_3l2}, the second row shows the acceptance after imposing the threshold cuts on the transverse momenta of the leptons in the two different categories $3l$ and $4l$, for the four BPs of Tab.~\ref{Tab:H1BPs}. The third row is the final acceptance after implementing all cuts plus efficiency, as in Tabs.~\ref{tab.ptleptoncuts} and \ref{tab:MC_H1} (for brevity we just report the last row). 

. The first setup is found to be more efficient. Moreover, from the comparison of the results in Tab.~\ref{tab:IT_H1_3l1} with those of Tab.~\ref{tab:IT_H1} it is clear that the number of events collected in the $3l$ category using the $3l$-trigger is compatible with that of the $2l$ category of the previous analysis for the BP1 and BP2, and bigger for the BP3 and BP4. Summing up the two separate categories, $3l$ and $4l$, the number of events more than doubles by employing the tri-lepton trigger of the first type ($P_T^{l_1} >$ 20 GeV, $P_T^{l_2} >$ 20 GeV and $P_T^{l_3} >$ 10 GeV) as compared to the di-lepton trigger. We obtain 46 events in the BP4 scenario and 22 in the BP3 one (with the di-lepton trigger we had 20 and 11 events, respectively).

This simple example highlights the importance of decreasing the thresholds of the $p_T$ cuts on displaced tracks produced by rather soft long-lived particles. Indeed, if implemented this would allow for the exploration of a new decay channel of the SM-like Higgs.

\section{Heavy neutrinos from the heavy Higgs}
\label{sec:H2}

In this section, we consider the region of the parameter space in which the long-lived heavy neutrinos have a mass greater than the threshold for their on-shell production coming from the 125 GeV Higgs boson decay. In this case, where $m_{\nu_h} > $ 62.5 GeV, the RH neutrinos can be produced from the decay of the heavy Higgs with a cross-section given by
\bea
\label{eq:sigmaxBR_H2}
\sigma(pp \rightarrow H_2 \rightarrow \nu_h \nu_h) = \frac{\sin^2 \alpha \, \sigma(pp \rightarrow H_2)_\textrm{SM} \Gamma( H_2 \rightarrow \nu_h \nu_h)}{\sin^2 \alpha \, \Gamma^\textrm{tot}_\textrm{SM}(H_2) + \Gamma( H_2 \rightarrow \nu_h \nu_h) +  \Gamma( H_2 \rightarrow H_1 H_1)} \,,
\eea
where $\sigma(pp \rightarrow H_2)_\textrm{SM}$ and $\Gamma^\textrm{tot}_\textrm{SM}(H_2)$ are the production cross section and  total width of a SM-like Higgs state with mass $m_{H_2}$, respectively, while $\Gamma( H_2 \rightarrow \nu_h \nu_h)$ and $\Gamma( H_2 \rightarrow H_1 H_1)$ are the partial decay widths of the heavy Higgs into two heavy neutrinos (summed over the three families) and two SM-like Higgs. They are given by
\bea
\Gamma( H_2 \rightarrow \nu_h \nu_h) &=& \frac{3}{2}  \frac{m_{\nu_h}^2}{x^2} \cos^2 \alpha \frac{m_{H_2}}{8 \pi} \left( 1 -  \frac{4 m_{\nu_h}^2}{m_{H_2}^2} \right)^{3/2} \,, \nonumber \\
 \Gamma( H_2 \rightarrow H_1 H_1) &=& \left( \frac{ \sin 2 \alpha}{v \, x} (x \cos \alpha + v \sin \alpha)  (\frac{m_{H_2}^2}{2} + m_{H_1}^2) \right)^2 \frac{1}{32 \pi \, m_{H_2}} \left( 1 -  \frac{4 m_{\nu_h}^2}{m_{H_2}^2} \right)^{1/2} 
\eea
Notice that the production cross section mediated by the heavy Higgs in Eq.~(\ref{eq:sigmaxBR_H2}) is similar to that in Eq.~(\ref{eq:sigmaxBR}), where the light Higgs is involved, the only difference is its dependence on the complementary scalar mixing angle and the appearance of the extra decay mode $H_2 \rightarrow H_1 H_1$. 

\begin{figure}
\centering
\subfigure[]{\includegraphics[scale=0.39]{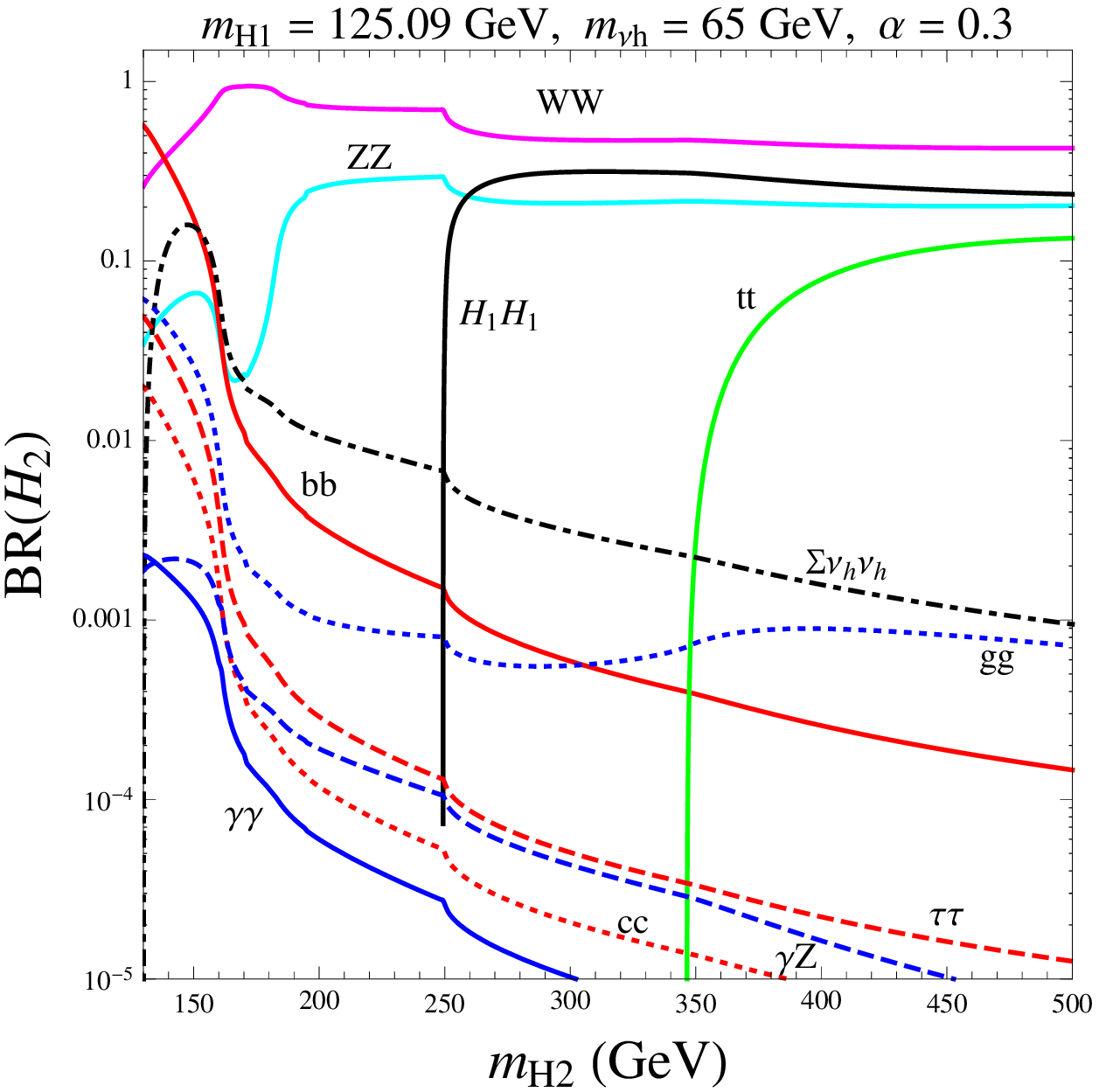}} 
\subfigure[]{\includegraphics[scale=0.39]{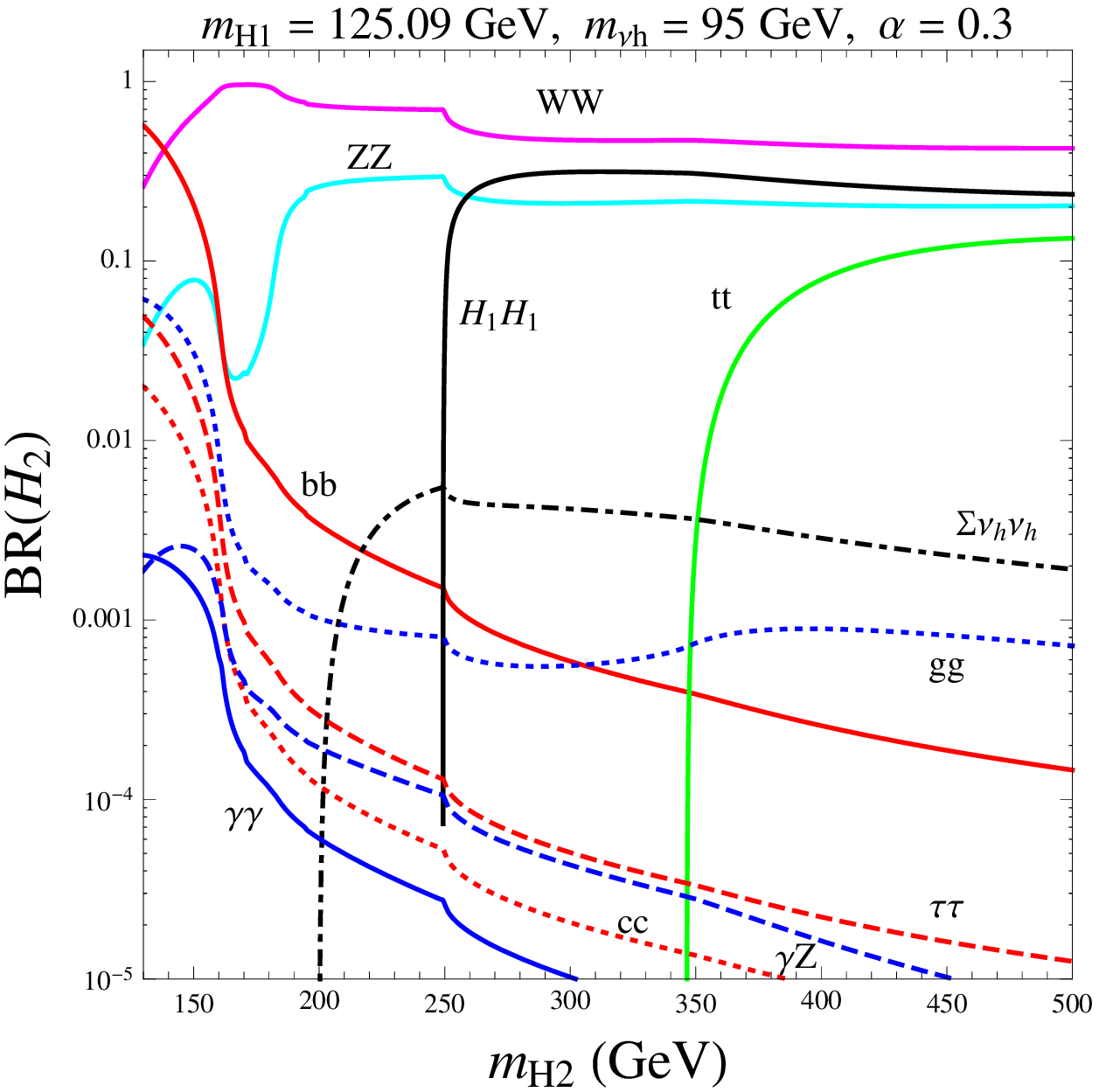}} 
\subfigure[]{\includegraphics[scale=0.54]{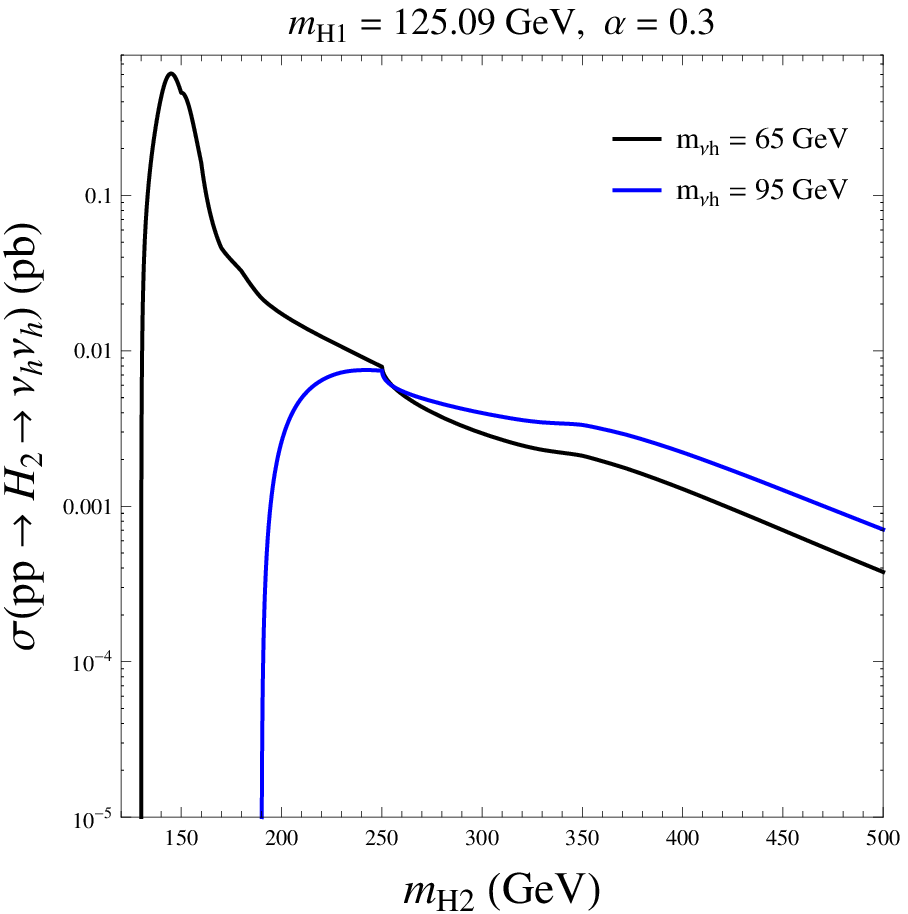}} 
\caption{Branching ratios of the heavy Higgs $H_2$ for (a) $m_{\nu_h} = 65$ GeV and (b) $m_{\nu_h} = 95$ GeV.  (c) Heavy neutrino pair production rate as a function of the heavy Higgs mass for two values of the heavy neutrino mass: $m_{\nu_h}$ = 65 GeV (black line) and $m_{\nu_h}$ = 95 GeV (blue line) .\label{Fig.BRH2}}
\end{figure}

We illustrate in Fig.~\ref{Fig.BRH2} the BRs of the heavy Higgs as a function of its mass $m_{H_2}$ for two values of the heavy neutrino mass, namely, (a) $m_{\nu_h} = 65$ GeV and (b) $m_{\nu_h} = 95$ GeV. In both cases, the decay mode $H_1 \rightarrow \nu_h \nu_h$ is kinematically closed and the heavy neutrino production from $H_2$ becomes the leading production mechanism (the corresponding cross section mediated by the $Z'$ for the same BP is $\sim 0.8$ fb). For $m_{\nu_h} = 65$ GeV, the $\textrm{BR}(H_2\rightarrow \nu_h\nu_h)$ can reach 10\% in the low $m_{H_2}$ region in which the on-shell decay modes into $WW$ and $ZZ$ are forbidden. In Fig.~\ref{Fig.BRH2}(c) we plot the heavy neutrino production cross section through $H_2$ for the same heavy neutrino masses discussed above showing that it can reach 456.8 fb for $m_{\nu_h} = 65$ GeV and $m_{H_2} = 150$ GeV. 

We repeat the analysis presented in the previous sections for the two BPs given in Tab.~\ref{Tab:H2BPs}. Due to the heavy neutrino masses being bigger than those of the BPs in the $H_1$-mediated case, the corresponding proper decay lengths are found to be smaller, thus pointing to the analysis of displaced leptons in the inner tracker as the most sensitive. The results are shown in Tabs.~\ref{tab:MC_H2} and \ref{tab:IT_H2} for the study of DVs and hits in the muon chambers and inner tracker, respectively. As expected, we count a larger number of events in both  BPs if we reconstruct the leptons using the information acquired from the tracker. Moreover, the number of events with displaced objects identified in BP6 is clearly much smaller than in BP5 due to the very different cross sections. Nevertheless, it is worth noticing that the overall efficiency in the reconstruction of the displaced tracks is a factor of $\sim 2$ greater in the BP6 case when the analysis in the tracker is concerned because its shorter $c \tau_0$ favours heavy neutrinos decaying in the inner part of the detector. 

Summing over the three disjoint categories, the number of expected events after reconstruction and cuts is around 223 within the BP5 scenario, characterised by a rather low $H_2$ mass. The BP6 scenario is less sensitive, providing a sample of barely 10 events. Interestingly, in the BP5 case the $4l$ category could offer the possibility of reconstructing a visible mass distribution that could give information about $m_{H_2}$. The signal sample is rich enough, with 17 events and a negligibly small background contribution. 

Finally, the flavour compositions of the events is shown in Tabs.~\ref{tab:IT_H2_2-3flavour} and \ref{tab:IT_H2_4flavour}.

\begin{table}
\centering
\begin{tabular}{|c||cc|c|c||c|}
\hline 
 	& $m_{H_2} \, \textrm{(GeV)}$ &  $m_{\nu_h} \, \textrm{(GeV)}$ & $m_{\nu_l} \, \textrm{(eV)}$ & $c \tau_0 \, \textrm{(m)}$ & $\sigma_{\nu_h\nu_h} \, \textrm{(fb)}$ \\ \hline
   BP5 & 150 & 65 & 0.023 & 0.5 &  456.8 \\  
   BP6 & 250 & 95 & 0.001 & 0.15 &  7.45 \\  \hline
\end{tabular}
\caption{Definition of the BPs leading to DVs from the heavy neutrinos produced from the heavy Higgs.  From left to right, the columns display the values of heavy Higgs mass, heavy neutrino mass, light neutrino mass, heavy neutrino proper decay length and heavy neutrino pair production rate.\label{Tab:H2BPs}}
\end{table}

\begin{table}
\centering
\begin{sideways}
\begin{tabular}{|c||c|c|c|c|c|c|c|c|c|c|c|c|c|c|c|}
\hline
\backslashbox{\scriptsize $p_T^{(1)}$/GeV}{\scriptsize $p_T^{(2)}$/GeV}   & 0 & 1 & 3 & 5 & 7 & 9 & 11 & 13 & 15 & 17 & 19 & 21 & 23 & 25 & 26\\ \hline \hline
0 & 100 &  &  &  &  &  &  &  &  &  &  &  &  &  & \\ \hline
1 & 96.53 & 94.31 &  &  &  &  &  &  &  &  &  &  &  &  & \\ \hline
3 & 87.26 & 85.41 & 79.77 &  &  &  &  &  &  &  &  &  &  &  & \\ \hline
5 & 75.78 & 74.3 & 69.75 & 63.69 &  &  &  &  &  &  &  &  &  &  & \\ \hline
7 & 63.35 & 62.2 & 58.66 & 53.92 & 48.47 &  &  &  &  &  &  &  &  &  & \\ \hline
9 & 51.75 & 50.87 & 48.17 & 44.52 & 40.32 & 36.27 &  &  &  &  &  &  &  &  & \\ \hline
11 & 41.65 & 40.97 & 38.9 & 36.12 & 32.93 & 29.83 & 26.92 &  &  &  &  &  &  &  & \\ \hline
13 & 33.05 & 32.53 & 30.95 & 28.83 & 26.42 & 24.11 & 21.9 & 19.85 &  &  &  &  &  &  & \\ \hline
15 & 25.83 & 25.43 & 24.23 & 22.61 & 20.8 & 19.07 & 17.44 & 15.89 & 14.44 &  &  &  &  &  & \\ \hline
17 & 19.82 & 19.51 & 18.6 & 17.4 & 16.07 & 14.8 & 13.59 & 12.46 & 11.39 & 10.44 &  &  &  &  & \\ \hline
19 & 14.95 & 14.72 & 14.04 & 13.14 & 12.18 & 11.25 & 10.36 & 9.534 & 8.754 & 8.079 & 7.454 &  &  &  & \\ \hline
21 & 11.06 & 10.89 & 10.38 & 9.731 & 9.028 & 8.352 & 7.721 & 7.126 & 6.577 & 6.108 & 5.672 & 5.304 &  &  & \\ \hline
23 & 8.039 & 7.914 & 7.538 & 7.068 & 6.553 & 6.066 & 5.616 & 5.195 & 4.807 & 4.485 & 4.193 & 3.956 & 3.756 &  & \\ \hline
25 & 5.808 & 5.719 & 5.442 & 5.095 & 4.721 & 4.37 & 4.046 & 3.75 & 3.483 & 3.264 & 3.069 & 2.922 & 2.796 & 2.709 & \\ \hline
26 & 4.907 & 4.83 & 4.591 & 4.289 & 3.973 & 3.677 & 3.406 & 3.161 & 2.942 & 2.764 & 2.605 & 2.487 & 2.389 & 2.327 & 2.304\\ \hline
\end{tabular} 
\end{sideways}
\caption{Efficiencies (\%) of the combined cuts $p_T^{lead.~l} > p_T^{(1)}$ on the two most energetic leptons and $p_T^{sublead.~l} > p_T^{(2)}$ on the third lepton for BP1 in Tab.~\ref{Tab:H1BPs}. \label{tab.ptleptoncuts}}
\end{table}

\begin{sidewaystable}
\begin{tabular}{|c|ccc|ccc|ccc|ccc|}
\hline
& \multicolumn{3}{c|}{BP1} & \multicolumn{3}{c|}{BP2} & \multicolumn{3}{c|}{BP3} & \multicolumn{3}{c|}{BP4} \\
					& $2\mu$ & $3 \mu$ & $4\mu$ & $2\mu$ & $3 \mu$ & $4\mu$ & $2\mu$ & $3 \mu$ & $4\mu$ & $2\mu$ & $3 \mu$ & $4\mu$ \\
\hline
Ev. before cuts 	 & 5016 & 960.2 & 57.57 &  3771 & 756.7 & 47.65 &  4699 & 922.9 & 55.64 &  3771 & 756.7 & 47.65 \\
$p_T$ cuts & 206.7 & 47.37 & 3.084 &  112.6 & 29.47 & 1.67 &  159.9 & 39.15 & 2.562 &  112.6 & 29.47 & 1.67 \\
$|\eta| < 2$ & 149.4 & 32.59 & 1.965 &  83.7 & 20.86 & 1.183 &  118.8 & 27.13 & 1.749 &  83.7 & 20.86 & 1.183 \\
$\Delta R > 0.2$ & 147.8 & 28.42 & 1.542 &  82.3 & 17.94 & 0.9191 &  117.5 & 23.43 & 1.395 &  82.3 & 17.94 & 0.9191 \\
$\cos \theta_{\mu\mu} > - 0.75$ & 114 & 19.33 & 0.9453 &  64.54 & 12.96 & 0.5839 &  92.7 & 15.86 & 0.9211 &  64.54 & 12.96 & 0.5839 \\
$L_{xy} < 5$ m & 100.7 & 17.59 & 0.8279 &  55.52 & 11.42 & 0.5408 &  89.31 & 15.47 & 0.907 &  62.63 & 12.77 & 0.5839 \\
$L_{xy}/\sigma_{L_{xy}} > 12$ & 63.19 & 10.62 & 0.6247 &  32.09 & 6.549 & 0.292 &  40.49 & 7.238 & 0.4335 &  10.91 & 2.068 & 0.06476 \\
$|L_z| < 8$ m & 53.97 & 8.717 & 0.5086 &  26.86 & 5.717 & 0.2056 &  37.1 & 6.71 & 0.3924 &  10.69 & 2.023 & 0.06476 \\
$|d_0|/\sigma_d > 4$ & 36.46 & 5.363 & 0.2764 &  19.6 & 3.847 & 0.1727 &  24.03 & 3.923 & 0.1351 &  6.201 & 0.9077 & 0.02159 \\
rec. eff. & 29.53 & 3.909 & 0.1813 &  15.88 & 2.804 & 0.1133 &  19.46 & 2.86 & 0.08865 &  5.023 & 0.6617 & 0.01416 \\
\hline
\end{tabular}
\caption{Displaced muons in the muon chambers originated from heavy neutrinos produced by the SM-like Higgs. For each BP the initial number of events generated is the cross section $\sigma(pp \rightarrow H_1 \rightarrow \nu_h \nu_h)$ multiplied by a luminosity of 100 fb$^{-1}$. The $p_T$ cuts are applied in the following way: $p_T > 26$ GeV for the two most energetic muons and $p_T > 5$ GeV for all the others. The reconstruction efficiency of the muons has taken into account. \label{tab:MC_H1}}
\bigskip\bigskip
\begin{tabular}{|c|ccc|ccc|ccc|ccc|}
\hline
& \multicolumn{3}{c|}{BP1} & \multicolumn{3}{c|}{BP2} & \multicolumn{3}{c|}{BP3} & \multicolumn{3}{c|}{BP4} \\
					& $2l$ & $3 l$ & $4l$ & $2l$ & $3 l$ & $4l$ & $2l$ & $3 l$ & $4l$ & $2l$ & $3 l$ & $4l$ \\
\hline
Ev. before cuts & 8891 & 4268 & 816.5 &  6645 & 3285 & 645.2 &  8307 & 4052 & 780.9 &  6645 & 3285 & 645.2  \\
$p_T$ cuts & 374 & 243.7 & 43.49 &  206.7 & 145.5 & 26.87 &  295 & 208.5 & 36.63 &  206.7 & 145.5 & 26.87 \\
$|\eta| < 2$ & 271 & 164.4 & 27.79 &  153.8 & 99.68 & 17.51 &  215.9 & 140.9 & 23.8 &  153.8 & 99.68 & 17.51 \\
$\Delta R > 0.2$ & 266.1 & 145.3 & 22.66 &  148.5 & 86.45 & 13.98 &  211.4 & 124.5 & 19.31 &  148.5 & 86.45 & 13.98 \\
$\cos \theta_{\mu\mu} > - 0.75$ & 239.4 & 129.3 & 19.34 &  135 & 78.52 & 12.22 &  192.1 & 112.5 & 16.59 &  135 & 78.52 & 12.22 \\
$10 < |L_{xy}| < 50$ cm & 13.24 & 6.798 & 1.265 &  9.259 & 5.082 & 0.7555 &  25.11 & 15.74 & 2.232 &  46.41 & 27.35 & 4.474 \\
$|L_{z}| < 1.4$ & 12.09 & 6.391 & 1.236 &  8.027 & 4.358 & 0.7339 &  22.38 & 14.61 & 2.043 &  41.51 & 25.17 & 4.29 \\
$|d_0|/\sigma_d > 12$ & 11.91 & 6.362 & 1.207 &  8.006 & 4.272 & 0.7339 &  22.19 & 14.39 & 1.962 &  40.94 & 24.96 & 4.247 \\
rec. eff. & 9.65 & 4.638 & 0.7916 &  6.485 & 3.114 & 0.4815 &  17.98 & 10.49 & 1.287 &  33.16 & 18.2 & 2.786 \\
\hline
\end{tabular}
\caption{Displaced leptons in the inner tracker originated from heavy neutrinos produced by the SM-like Higgs. For each BP the initial number of events generated is the cross section $\sigma(pp \rightarrow H_1 \rightarrow \nu_h \nu_h)$ multiplied by a luminosity of 100 fb$^{-1}$. The $p_T$ cuts are applied in the following way: $p_T > 26$ GeV for the two most energetic leptons and $p_T > 5$ GeV for all the others. The reconstruction efficiency of the leptons has taken into account. \label{tab:IT_H1}}
\end{sidewaystable}

\begin{table}
\centering
\begin{tabular}{|c||c|c|c||c|c|c|}
\hline
			&  $ee$ 	&  $\mu\mu$ 	&  $e \mu$ 		&  $(ee)l$ 	&  $(\mu\mu)l$ 	&  $(e \mu)l$  \\ \hline
 BP1 	& 4.943 & 3.859 & 0.8485		& 1.154 & 0.6992 & 2.785		\\
 BP2 	& 3.361 & 2.581 & 0.5429		& 0.67 & 0.6062 & 1.838		\\ 
 BP3 	& 9.557 & 7.179 & 1.24		& 2.9 & 1.589 & 6.002		\\
 BP4 	& 17.09 & 13.09 & 2.978		& 3.874 & 3.293 & 11.03	\\ \hline
\end{tabular}
\caption{Flavour composition of 2- and 3-displaced leptons in the tracker. \label{tab:IT_H1_2-3flavour}}
\bigskip \bigskip
\centering
\begin{tabular}{|c||c|c|c|c|c|c|}
\hline
			&  $(ee)(ee)$ 	&  $(ee)(\mu\mu)$ 	&  $(\mu\mu)(\mu\mu)$ 		&  $(ee)(e\mu)$ 	&  $(\mu\mu)(e\mu)$ 	&  $(e \mu)(e\mu)$  \\ \hline
 BP1 	& 0.03809 & 0.05714 & 0 & 0.2103 & 0.1905 & 0.2956		\\
 BP2 	& 0.07081 & 0.04249 & 0.01416 & 0.1275 & 0.1133 & 0.1133		\\ 
 BP3 	& 0.06243 & 0.06243 & 0.05319 & 0.4432 & 0.1596 & 0.5064		\\
 BP4 	& 0.2124 & 0.1774 & 0.1491 & 0.7089 & 0.6023 & 0.9362		\\ \hline
\end{tabular}
\caption{Flavour composition of 4-displaced leptons in the tracker. \label{tab:IT_H1_4flavour}}
\end{table}

\begin{table}
\centering
\begin{tabular}{|c|cc|cc|cc|cc|}
\hline
& \multicolumn{2}{c|}{BP1} & \multicolumn{2}{c|}{BP2} & \multicolumn{2}{c|}{BP3} & \multicolumn{2}{c|}{BP4} \\
					 & $3 l$ & $4l$  & $3 l$ & $4l$  & $3 l$ & $4l$  & $3 l$ & $4l$ \\
\hline
Ev. before cuts & 4268 & 816.5 &  3285 & 645.2  & 4052 & 780.9  & 3285 & 645.2 \\
$p_T$ cuts & 417.1 & 105 &   289.5 & 79.46  & 369.6 & 97.07  & 289.5 & 79.46 \\
rec. eff. & 8.218 & 1.421  & 7.122 & 1.362  & 19.33 & 3.471  & 38.61 & 8.105 \\
\hline
\end{tabular}
\caption{Displaced leptons originated from heavy neutrinos produced by the SM-like Higgs.  For each BP the initial number of events generated is the cross section $\sigma(pp \rightarrow H_1 \rightarrow \nu_h \nu_h)$ multiplied by a luminosity of 100 fb$^{-1}$. A $3l$-trigger has been employed. The $p_T$ cuts are applied in the following way: $p_T > 20$ GeV for the two most energetic leptons and $p_T > 10$ GeV for the third one. A $p_T > 5$ GeV is required on the forth lepton, when present. Only the events surviving the final selection procedure and the cuts on the $p_T$ are shown. \label{tab:IT_H1_3l1}}
\bigskip \bigskip
\begin{tabular}{|c|cc|cc|cc|cc|}
\hline
& \multicolumn{2}{c|}{BP1} & \multicolumn{2}{c|}{BP2} & \multicolumn{2}{c|}{BP3} & \multicolumn{2}{c|}{BP4} \\
					 & $3 l$ & $4l$  & $3 l$ & $4l$  & $3 l$ & $4l$  & $3 l$ & $4l$ \\
\hline
Ev. before cuts  & 4268 & 816.5 & 3285 & 645.2 & 4052 & 780.9  & 3285 & 645.2 \\
$p_T$ cuts & 290 & 86.03 & 201.5 & 65.81 & 260.4 & 79.96 & 201.5 & 65.81 \\
rec. eff.  & 5.307 & 1.305  & 5.608 & 1.369  & 13.98 & 3.425  & 27.29 & 6.942 \\
\hline
\end{tabular}
\caption{Displaced leptons originated from heavy neutrinos produced by the SM-like Higgs.  For each BP the initial number of events generated is the cross section $\sigma(pp \rightarrow H_1 \rightarrow \nu_h \nu_h)$ multiplied by a luminosity of 100 fb$^{-1}$. A $3l$-trigger has been employed. The $p_T$ cuts are applied in the following way: $p_T > 20$ GeV for the most energetic lepton and $p_T > 15$ GeV for the other two. A $p_T > 5$ GeV is required on the forth lepton, when present. Only the events surviving the final selection procedure and the cuts on the $p_T$ are shown. \label{tab:IT_H1_3l2}}
\end{table}

\begin{table}
\centering
\begin{tabular}{|c|ccc|ccc|}
\hline
& \multicolumn{3}{c|}{BP5} & \multicolumn{3}{c|}{BP6} \\
					& $2\mu$ & $3 \mu$ & $4\mu$ & $2\mu$ & $3 \mu$ & $4\mu$ \\
\hline
Ev. before cuts & 7266 & 1589 & 108 &  162.8 & 63.69 & 7.519 \\
$p_T$ cuts & 323.7 & 109.8 & 10.12 &  19.38 & 13.22 & 3.27 \\
$|\eta| < 2$ & 242 & 74.59 & 6.479 &  16.18 & 10.29 & 2.316 \\
$\Delta R > 0.2$ & 240.8 & 68.04 & 5.411 &  15.22 & 6.217 & 1.377 \\
$\cos \theta_{\mu\mu} > - 0.75$ & 207.9 & 51.68 & 3.763 &  8.913 & 4.316 & 0.8433 \\
$L_{xy} < 5$ m & 206.3 & 51.57 & 3.763 &  8.54 & 4.298 & 0.8433 \\
$L_{xy}/\sigma_{L_{xy}} > 12$ & 14.9 & 3.252 & 0.2598 &  1.642 & 0.238 & 0.006817 \\
$|L_z| < 8$ m & 14.42 & 3.252 & 0.2598 &  1.608 & 0.233 & 0.006817 \\
$|d_0|/\sigma_d > 4$ & 9.9 & 2.049 & 0.2369 &  0.3646 & 0.05198 & 0.000554 \\
rec. eff. & 8.019 & 1.494 & 0.1554 &  0.2953 & 0.03789 & 0.0003634 \\
\hline
\end{tabular}
\caption{Displaced muons in the muon chambers originated from heavy neutrinos produced by the heavy Higgs.  For each BP the initial number of events generated is the cross section $\sigma(pp \rightarrow H_2 \rightarrow \nu_h \nu_h)$ multiplied by a luminosity of 100 fb$^{-1}$. The $p_T$ cuts are applied in the following way: $p_T > 26$ GeV for the two most energetic muons and $p_T > 5$ GeV for all the others. The reconstruction efficiency of the muons is  taken into account. \label{tab:MC_H2}}
\end{table}
%
%
%
%
%
\begin{table}
\centering
\begin{tabular}{|c|ccc|ccc|}
\hline
& \multicolumn{3}{c|}{BP5} & \multicolumn{3}{c|}{BP6} \\
					& $2l$ & $3 l$ & $4l$ & $2l$ & $3 l$ & $4l$ \\
\hline
Ev. before cuts & 12556 & 6612 & 1371 &  224.9 & 202.4 & 64.88 \\
$p_T$ cuts & 553.3 & 531.7 & 125.3 &  31.17 & 44.95 & 27.99 \\
$|\eta| < 2$ & 414.6 & 361.7 & 78.13 &  25.89 & 34.81 & 19.57 \\
$\Delta R > 0.2$ & 409.9 & 333.7 & 67.6 &  22.92 & 21.43 & 11.82 \\
$\cos \theta_{\mu\mu} > - 0.75$ & 385.3 & 306.9 & 60.81 &  19.63 & 18.8 & 10 \\
$10 < |L_{xy}| < 50$ cm & 174.8 & 135.5 & 28.49 &  5.327 & 6.596 & 3.647 \\
$|L_{z}| < 1.4$ & 149.3 & 118.9 & 26.31 &  5.202 & 6.428 & 3.569 \\
$|d_0|/\sigma_d > 12$ & 147.9 & 117.5 & 26.11 &  5.088 & 6.278 & 3.462 \\
rec. eff. & 119.8 & 85.63 & 17.13 &  4.121 & 4.577 & 2.272 \\
\hline
\end{tabular}
\caption{Displaced leptons originated from heavy neutrinos produced by the heavy Higgs.  For each BP the initial number of events generated is the cross section $\sigma(pp \rightarrow H_2 \rightarrow \nu_h \nu_h)$ multiplied by a luminosity of 100 fb$^{-1}$. The $p_T$ cuts are applied in the following way: $p_T > 26$ GeV for the two most energetic muons and $p_T > 5$ GeV for all the others. The reconstruction efficiency of the leptons is taken into account. \label{tab:IT_H2}}
\end{table}

\begin{table}
\centering
\begin{tabular}{|c||c|c|c||c|c|c|}
\hline
			&  $ee$ 	&  $\mu\mu$ 	&  $e \mu$ 		&  $(ee)l$ 	&  $(\mu\mu)l$ 	&  $(e \mu)l$  \\ \hline
 BP5		& 55.55 & 49.69 & 14.55		& 19.13 & 16.29 & 50.21	\\
 BP6		& 1.457 & 1.111 & 1.553		& 1.31 & 0.9903 & 2.277  \\ \hline
\end{tabular}
\caption{Flavour composition of 2- and 3-displaced leptons in the tracker. \label{tab:IT_H2_2-3flavour}}
\bigskip\bigskip
\begin{tabular}{|c||c|c|c|c|c|c|}
\hline
			&  $(ee)(ee)$ 	&  $(ee)(\mu\mu)$ 	&  $(\mu\mu)(\mu\mu)$ 		&  $(ee)(e\mu)$ 	&  $(\mu\mu)(e\mu)$ 	&  $(e \mu)(e\mu)$  \\ \hline
 BP5		& 1.243 & 0.8808 & 0.9517 & 3.998 & 3.639 & 6.419 \\
 BP6		& 0.3146 & 0.03244 & 0.1942 & 0.5922 & 0.5579 & 0.5802 \\ \hline
\end{tabular}
\caption{Flavour composition of 4-displaced leptons in the tracker. \label{tab:IT_H2_4flavour}}
\end{table}

\section{Conclusions}
\label{sec:summa}
In summary, we have assessed the significant scope that Run 2 of the LHC can have in exploring the possibility of the existence of new neutrinos, heavier than the SM ones, yet with a mass below the EW scale (100 GeV or so). These objects are almost ubiquitous in the class of  BSM scenarios aimed at addressing the puzzle that emerged from the discovery of SM neutrino flavour oscillations, hence and the need to explain their masses. 

Furthermore, such states, owing to their EW interactive nature (i.e., with small coupling strengths), relative lightness (so that the phase space open to their decays is small) and the fact that the decay currents proceed via off-shell weak gauge bosons, are generally long lived. In fact, for a significant portion of the parameter space of the BSM scenarios hosting them,
they could often decay inside an LHC detector. Depending on the actual decay length, they can do so and be identified (given the SM lepton flavour they generate while decaying) in the inner tracking system or the muon chambers
(emulated here through the CMS parameters, though simple adaptations to the ATLAS environment can equally be pursued).
In either case, one or two DVs would be visible against a negligibly small background environment, based on well-established triggers available for the CMS detector. Indeed, we have also highlighted the importance that the exploitation of new triggers, specifically, displaced tri-lepton ones, could have for this DV search.    

Among st the possible production modes of such heavy neutrino states, we have concentrated here on the case
of both light (i.e., SM-like) and heavy (i.e., via an additional state) Higgs mediation starting from gluon-gluon fusion. On the one hand, this approach complements earlier analyses based on $Z'$ mediation, which is now greatly suppressed in the light of the latest experimental limits on $Z'$ masses. On the other hand, it also offers sensitivity (indeed at standard energies and luminosities foreseen for Run 2 of the LHC) to the aforementioned class of BSM scenarios.

In this connection, it should finally be noted that, while we have benchmarked our analysis against the specific realisation of the class of anomaly-free, non-exotic, minimal $U(1)'$ extensions of the SM with a specific $\tilde g/g^\prime_1$ ratio (for the purpose of being quantitative), our conclusions can generally be applied to the whole category of such models. 

In short, we believe to have opened a rather simple path to follow towards the efficient exploration of new physics
scenarios remedying a significant flow of the SM, i.e., the absence of massive neutrinos.

\section*{Acknowledgements}
LDR thanks L. Basso for useful discussions during the development of this manuscript. We are also grateful to J. Fiaschi for extracting the experimental bounds on mass and couplings of the $Z'$-boson from the data analysis at the LHC RunII. Moreover, a great acknowledgment should go to Ian Tomalin for his guidance on the trigger thresholds.
EA, SM and CHS-T are supported in part through the NExT Institute. The work of LDR
has been supported by the ``Angelo Della Riccia'' foundation and the STFC/COFUND Rutherford International Fellowship
scheme.

\newpage

\providecommand{\href}[2]{#2}\begingroup\raggedright\endgroup

\end{document}